\documentclass[aps,twocolumn,amsfonts,amssymb,longbibliography,superscriptaddress]{revtex4-1}
\usepackage{graphicx}
\usepackage{multirow}
\usepackage{booktabs}
\usepackage{epstopdf} %converting to PDF
\usepackage{dcolumn}
\usepackage{bm}
\usepackage{bbm,bbold}
\usepackage{color}
\usepackage{xcolor, soul}
\sethlcolor{green}
\usepackage{amsfonts}
\usepackage{amsmath}
\usepackage{mdframed}
\usepackage[normalem]{ulem}
\usepackage{mathrsfs}   
\usepackage[none]{hyphenat}
\usepackage{subfigure}
\usepackage{float}
\usepackage{physics}
\usepackage[colorlinks=true,citecolor=blue]{hyperref}
\hypersetup{colorlinks=true,citecolor=blue,linkcolor=blue,urlcolor=blue}

\usepackage{makecell}
\usepackage{pifont}
\usepackage{tabularx}
\begin{document}
	\title{Unveiling Novel Resonant Interband Contribution to Polarizability in three-dimensional systems}	
	\author{Vivek Pandey}
    \email{vivek_pandey@srmap.edu.in}
	\affiliation{Department of Physics, School of Engineering and Sciences, SRM University AP, Amaravati, 522240, India}
    \author{Snehasish Nandy}
    \email{snehasish@phy.nits.ac.in}
	\affiliation{Department of Physics, National Institute of Technology, Silchar, Assam, 788010, India}
 	\author{Pankaj Bhalla}
	\email{pankaj.b@srmap.edu.in}
 \affiliation{Department of Physics, School of Engineering and Sciences, SRM University AP, Amaravati, 522240, India}
\affiliation{Centre for Computational and Integrative Sciences, SRM University AP, Amaravati, 522240, India}
    
\date{\today}

\begin{abstract}
Polarizability plays an essential role in characterizing key phenomena, such as the screening effects, collective excitations, and dielectric functions present in the system. In three-dimensional materials, it typically comprises an intraband contribution, dependent on the chemical potential, and an interband contribution, largely independent of it. In this study, within the random phase approximation framework, we uncover a novel interband contribution that, unlike the conventional case, exhibits an explicit dependence on the chemical potential, which has no counterpart in two dimensions. In the long-wavelength limit, this term introduces a resonance feature with cubic wave-vector dependence when the chemical potential approaches the band edge, in contrast to the quadratic behavior characteristic of standard intraband and interband processes. Focusing on three-dimensional Dirac nodal line semimetals, we show that the polarizability is intraband-dominated at low frequencies, while interband processes prevail at intermediate and high frequencies, with the overall response being tunable via the chemical potential. Material-specific estimates for Ca$_3$P$_2$ and ZrSiS reveal a strong tunability of both contributions. These findings open new directions for probing frequency-dependent dielectric properties and hold promise for applications in tunable plasmonic and optoelectronic devices.

\end{abstract}

\maketitle

\section{Introduction}

Screening of Coulomb interactions arising from many-body effects plays a central role in determining diverse physical properties, and is effectively characterized by the polarizability and the dielectric function. The polarizability, or density-density response function, $\text{P}(\bm q, \omega)$ quantifies the system's response to a perturbing scalar electric potential $\phi (\bm q, \omega)$ with wave vector $\bm q$ and frequency $\omega$. Within linear response theory, it is defined as~\cite{giuliani_cup2008}: 
\begin{align}
    \delta n (\bm q, \omega) = \text{P}(\bm q, \omega) \phi (\bm q, \omega),
\end{align}
where $\delta n (\bm q, \omega)$ stands for the charge density caused by the potential applied. In the static case ($\omega=0$), the polarizability describes the screening Coulomb potential by charged impurities, while for the dynamic case ($\omega \neq 0$), it governs various phenomena such as plasmon excitations, optical absorption. More broadly, polarizability serves as a central tool to probe electronic properties and examine the behavior of topological systems.
\begin{figure}[htp]
    \centering
    \includegraphics[width=7.5 cm]{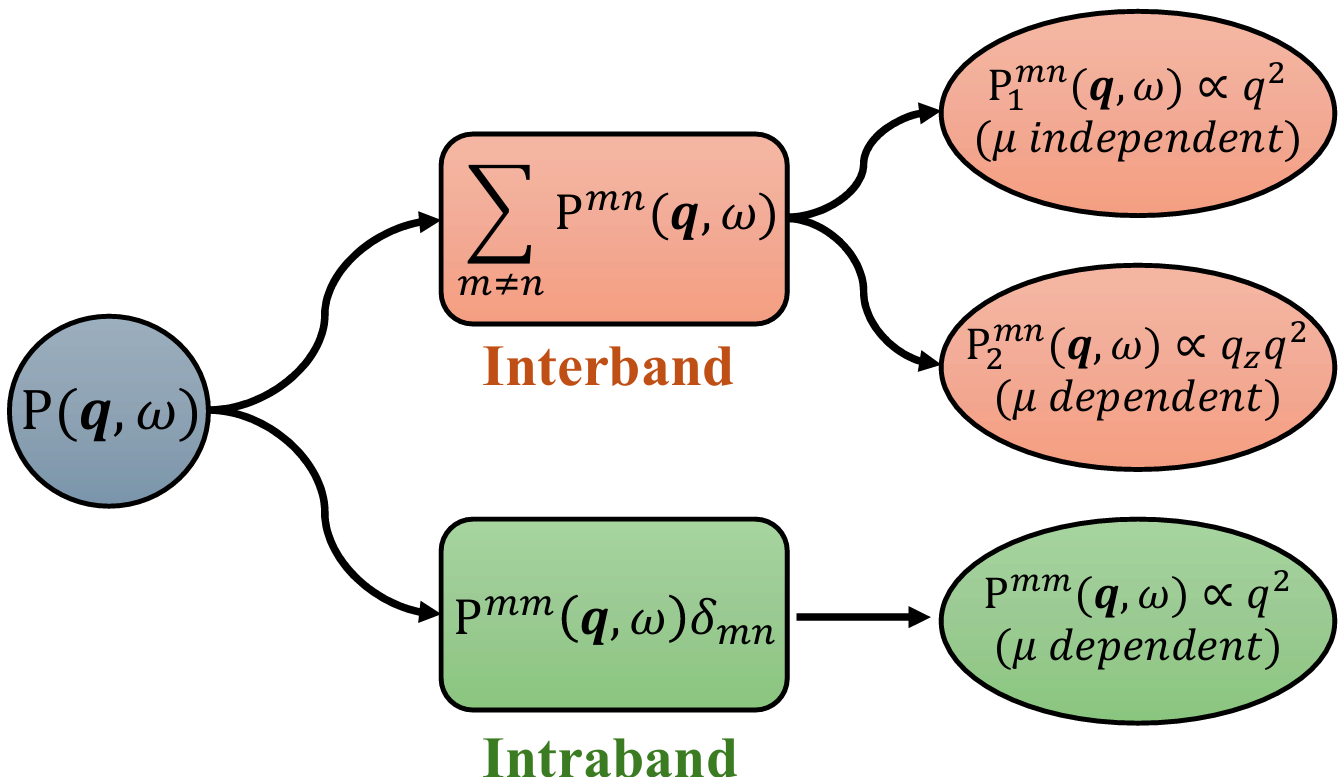}
    \caption{The schematic shows the different components of polarizability $\text{P}(\bm q, \omega)$, which consist of two main contributions: the interband part, $\sum\limits_{m\neq n}\text{P}^{mn}(\bm q, \omega)$, and the intraband part $\text{P}^{mm}(\bm q, \omega)\delta_{mn}$. Here, the wave vector is defined as $q=\sqrt{q_\rho^2+q_z^2}$, with $q_\rho = \sqrt{q_x^2 + q_y^2}$. Further, the interband polarizability comprises two parts, one is the $\mu$ independent and the other is the $\mu$ dependent part, which shows $q^2$ and $q_zq^2$ dependence, respectively. In contrast, the intraband part arises solely from $\mu$ dependent part.} 
    \label{fig:schematic}
\end{figure}

In light of these considerations, substantial efforts have been devoted to investigate polarizability in both two-dimensional (2D) and three-dimensional (3D) topological systems, revealing that it consists of two primary components: an intraband contribution, which is chemical potential ($\mu$) dependent and an interband contribution which is $\mu$ independent~\cite{giuliani_cup2008}. For example, in the case of graphene or surface states of 3D topological insulator~\cite{Hasan_rmp2010, hasan_arcmp2011, moore_n2010, Qi_rmp2011, franz_e2013}, upon doping, the dynamical polarizability shows linear dependence with $\mu$ and scales as $q^2/\omega$ at the low temperature and long wavelength limit~\cite{Dassarma_prb2015, Hwang_prb2007}. On the other hand, it is shown that the intrinsic contribution of static screening in this regime
originates from the interband transitions and can be effectively absorbed in a background dielectric constant~\cite{Hwang_prb2007}.

Going beyond 2D, in 3D Dirac~\cite{Armitage_rmp2018, Young_prb2012,liu_nm2014, Borisenko_prl2014, yi_sr2014,  Wang_prb2013, liu_science2014} and Weyl semimetals~\cite{Armitage_rmp2018, Yan_arcmp2017, Lv_prx2015, soluyanov_n2015, Hasan_arcmp2017, Weng_prx2015, xu_nc2016, Belopolski_prl2016, Xu_sa2015, Zyuzin_prb2012}, the dynamical polarizability exhibits $\mu^2$ and $q^2/\omega^2$ behavior driven by the intraband transitions~\cite{Dassarma_prb2015, Sarma_prl2009, Thakur_jpcm2017, wu_arxiv2018}. However, an additional contribution to the polarizability can arise in Weyl semimetals in the presence of non-orthogonal electric and magnetic fields due to the chiral anomaly~\cite{wu_arxiv2018}. In massive Dirac systems, the dynamical polarization function has been analyzed and finds a variation of screening potential as $r^{-n}$, where $r$ represents the spatial decay rate and $n$ refers to the dimensionality~\cite{Thakur_jpcm2017}. Collective modes have also been investigated in various systems, including multi-Weyl semimetals~\cite{ahn_sr2016} and a massless Dirac plasma~\cite{Sarma_prl2009}, where the interband transitions and chirality lead to a depolarization shift in plasma frequencies~\cite{ahn_sr2016}. 

In contrast to the aforementioned class of semimetallic systems, where the energy bands touch linearly at distinct points, a new class of topological semimetals, namely, nodal line semimetals (NLSMs), has emerged, characterized by band crossings in the form of rings or lines~\cite{Burkov_prb2011, Fang_cpb2016, Shuo_apx2018, Bian_prb2016, Fang_prb2015, Xie_APLM2015, Kim_prl2015, Ekahana_njp2017, Xu_prb2017, Takane_prb2017, Chen_prb2017, Wang_prb2017, chang_arxiv2025, Zhu_cpl2024}. These systems exhibit a complex polarizability behavior due to the nontrivial density of states \cite{Rukelj_s2024}. Specifically, the polarizability of NLSMs is proposed to be highly sensitive to factors such as the nodal ring radius, chemical potential, and gap term. In this context, Rahimpoor et al. and Yan et al. have analyzed collective modes and dielectric function in 2D and 3D NLSMs theoretically~\cite{Rahimpoor_prb2024, Yan_prb2016} whereas Xue et al. experimentally observed the plasmonic dispersion in ZrSiS system~\cite{Xue_prl2021}. It has been shown theoretically that for a tilted 2D nodal line semimetal, the static polarizability arises solely from the intraband transitions and remains isotropic despite of the electronic band structure anisotropy. The intrinsic (undoped case) polarizability depends on the nodal ring radius, whereas the extrinsic part depends on both the Fermi energy and the nodal ring radius. Moreover, in the long-wavelength limit, the plasmon dispersion follows a square root dependence on the wave vector~\cite{Rahimpoor_prb2024}. 
In contrast, the situation changes for 3D systems and for dynamical polarizability, where interband transitions cannot be neglected, as they significantly reshape the response and dominate in frequency regimes inaccessible to intraband contributions. Notably, Ref.~\cite{Yan_prb2016} demonstrates that, within the long-wavelength limit, the plasmon frequency in 3D NLSM receives contributions from both channels: the chemical potential dependent intraband part and the chemical potential independent interband part. Additionally, in ZrSiS, plasmons have been experimentally observed in both the intraband and interband regimes. The intraband plasmons stem from contributions of both the surface states and bulk nodal line state, whereas the interband plasmon arises from the transitions between the distinct surface states~\cite{Xue_prl2021}. 
%However, this work does not give the explicit dependence of these responses on the chemical potential.
%
Despite the above mentioned contributions, the emergence of the resonant interband transitions can also make a significant impact on the interband polarizability, which can arise from the surface state effects. Such transitions can be obtained in two ways: shifting the chemical potential to the edge of the conduction band and applying external energy equivalent to the gap between the conduction and valence bands. This particular aspect of the interband polarizability response has not been explored, motivating us to further in-depth exploration into the polarizability in 3D systems. 

In this work, we revisit the general expression for the polarizability within the random phase approximation (RPA) and uncover a novel interband contribution with cubic wave-vector dependence that, at low temperatures, acquires an explicit chemical-potential dependence. This term plays a pivotal role, as it generates a resonant signature in the response when the chemical potential approaches the band edge, providing a tunable optical fingerprint of the system. By analyzing the total polarizability in 3D Dirac nodal line semimetals using a low-energy two-band model, we identify that the leading-order interband response contains both $q^2$ and $q_z q^2$ terms. These originate from chemical-potential independent and dependent parts, respectively. Remarkably, the anisotropic $q_z q^2$ term, absent in the two-dimensional systems, exhibits a resonant behavior unique to 3D NLSMs and can even dominate over the conventional ($q^2$) doping-dependent contribution. This highlights the critical role of out-of-plane momentum transfer ($q_z$) in shaping the screening behavior of 3D NLSMs, in sharp contrast to their isotropic 2D counterparts. We further show that, while intraband processes dominate in the low-frequency regime, interband transitions become the leading contribution at intermediate and high frequency regimes. This dominance reflects strong band coupling near the nodal ring and fundamentally distinguishes the dynamical response of 3D NLSMs from that of 2D systems, where intraband channels prevail. We also explore the impact of a $\mathcal{PT}$-symmetry–breaking mass term. Besides opening a band gap and suppressing certain interband channels, the mass term induces a sign-changing behavior in the intraband response, suggestive of a metal–insulator transition. This establishes a direct connection between gap opening and dielectric response, which can be tuned experimentally via pressure, strain, or external electric fields. A schematic flow chart summarizing the different contributions to the total polarizability is presented in Fig.~\ref{fig:schematic}. The comparison between the literature and the present work is given in Table.~\ref{tab:1}. Finally, material-specific estimates for Ca$_3$P$_2$ and ZrSiS highlight the strong tunability of both intraband and interband components, for which we present detailed quantitative evaluations.
\begin{table}[t]
    \centering
    \fontsize{7pt}{7pt}\selectfont  % Compact font
    \renewcommand{\arraystretch}{1.5}  % Vertical spacing
    \caption{The table highlights the comparison between the literature and our work. Here $\text{P}_i^{mn}$ is the interband polarizability having the subscript $i=1, 2$ for chemical potential $\mu$-independent and dependent parts respectively, the superscript $m,n$ for band indices and $M$ represents the mass term.}
    \label{tab:1}
    \begin{ruledtabular}
    \begin{center}
    \setlength{\tabcolsep}{0pt}
    \begin{tabular}{ccc}
    
         & Literature~\cite{Rahimpoor_prb2024, Yan_prb2016, Xue_prl2021} & Our work \\
        \hline
        Polarizability 
        & $\text{P}_1^{mn}$ 
        & $\text{P}_1^{mn}+ \text{P}_2^{mn}$  \\
        Chemical potential dependence
        & \ding{55} & \ding{51} \\
        Resonance feature 
        & \ding{55} & \ding{51} $(\mu=M)$\\
        $q$-dependence 
        & $q^2$
        & $q^2,\,q_z q^2$ \\
    \end{tabular}
    \end{center}
    \end{ruledtabular}
\end{table}

\section{Different Contributions of Polarizability} 
\label{Sec:Polarizability and Model in Dirac nodal line semimetal}

In general, the polarizability of a system is defined through the density–density correlation function $\mathcal{P}(\bm q, t)=-i\Theta(t) \langle[\hat{n}_{\bm q}(t),\hat{n}_{-\bm q}(0)]\rangle$ with the density operator $\hat{n}_{\bm q} (t)=\sum_{n,\bm k}\hat{c}^\dagger_{n, \bm k + \bm q}(t)\hat{c}_{n, \bm k}(t)$ where $\hat{c}^\dagger_{n,\bm k+\bm q}(t)$ and $\hat{c}_{n,\bm k} (t)$ are the creation and annihilation operators associated with an electron having momentum $\bm k+\bm q$ and $\bm k$, $[\cdot,\cdot]$ represents the commutation bracket, $\langle \cdots\rangle$ denotes the thermal average and $\Theta(t)$ represents the step function. Here we consider $\hbar = 1$, thus treating the wave vector as momentum and frequency as energy in the same way throughout the calculations. By employing the time-ordering to ensure causality and Wick's theorem to simplify the product of four creation and annihilation operators, the polarizability in the frequency domain can be written as $\mathcal{P}(\bm q, \omega)=\sum\limits_{m,n}\text{P}^{mn}(\bm q, \omega)$, where~\cite{giuliani_cup2008, Hwang_prb2007, Yan_prb2016} 
\begin{align} \label{eqn: polarizability_gen}
    &\text{P}^{mn}(\bm q, \omega) = \frac{1}{2 v} \sum_{\bm k} \frac{f( \varepsilon^m_{\bm k+\bm q})-f(\varepsilon^n_{\bm k})}{\omega+ \varepsilon^m_{\bm k+\bm q}-\varepsilon^n_{\bm k}+i \eta} |\langle m, \bm k +\bm q|n, \bm k\rangle|^2.
\end{align}
Here $v$ corresponds to the volume of the system and is taken as unity for all calculations, $f(\varepsilon_{\bm k}^n)$ is the Fermi-Dirac distribution function that arises from the expectation value of the fermionic operators $\langle \hat{c}^\dagger_{m,\bm k+\bm q} \hat{c}_{n,\bm k}\rangle=\delta_{\bm k + \bm q, \bm k} \delta_{mn} f(\varepsilon_{\bm k + \bm q}^m)$, where $\varepsilon_{\bm k}$ is the energy dispersion and $m, n$ represent the band indices. Further, it is defined as $f(\varepsilon_{\bm k}^n)=[1+e^{\beta(\varepsilon_{\bm k}^{n}-\mu)}]^{-1}$ with $\beta=1/k_{B}T$ where $k_{B}$ is the Boltzmann constant, $T$ is the temperature associated with an electron and $ |\langle m, \bm k +\bm q|n, \bm k\rangle|^2$ is the transition probability,  lastly $\eta\rightarrow 0^+$ is a small positive infinitesimal number. 
%Additionally, $ |\langle m, \bm k +\bm q|n, \bm k\rangle|$ represents the matrix element between the electron's initial and final states, $ |\langle m, \bm k +\bm q|n, \bm k\rangle|^2$ is the transition probability, which is defined as $|\langle m, \bm k +\bm q|n, {\bm k} \rangle|^2= 1+ mn \cos{2 \delta \theta_{\bm{kq}}^{mn}}$, where $\delta\theta_{\bm{kq}}^{mn}=\theta_{\bm k +\bm q}^m-\theta_{\bm k}^n$ is the angle between $\bm k$ and $\bm k + \bm q$ and lastly $\eta\rightarrow 0^+$ is a small positive infinitesimal number.   

In the present study, we employ the long-wavelength limit $q \to 0$, which is the regime of primary interest and evaluate the real part of the polarizability, capturing the essential features of screening as well as the static and dynamical response of the system. The imaginary part, corresponding to absorption processes, vanishes for the model considered here in the specific regime discussed in Sec.~\ref{Sec:III}. In this limit, one can obtain a simplified expression of polarizability using the Taylor expansion with respect to $q$, which is given by (for detailed derivation, see Appendix~\ref{Appendix:A1})

%Further, one can simplify the Eq.~(\ref{eqn: polarizability_gen}) by employing the Taylor expansion of several factors with respect to $q$: (i) difference between the distribution functions $f( \varepsilon^m_{\bm k+\bm q})-f(\varepsilon^n_{\bm k})$, (ii) energy difference associated with a different momentum $ \varepsilon^m_{\bm k+\bm q}-\varepsilon^n_{\bm k}$ and (iii) angular factor $\cos{2\delta\theta^{mn}_{\bm{kq}}}$. In this regime, using the expansions of all above factors upto $\mathcal O(q^2)$, the form of the polarizability given in Eq.~\eqref{eqn: polarizability_gen} can be written as (for detailed derivation, see Appendix~\ref{Appendix:A1})

\begin{align} \label{eqn:polarizability_sim} \nonumber
&\text{P}^{mn}(\mathbf{q}\rightarrow 0, \omega) = \\ \nonumber &\frac{1}{2} \sum_{\bm k} 
\frac{f^{mn}+ \sum_\gamma q_\gamma \partial_{\gamma} f(\varepsilon_{\bm k}^m)+\frac{1}{2}\sum_{\alpha, \beta}q_\alpha q_\beta\,\partial_{\alpha\beta} f(\varepsilon^m_{\bm k})}{\omega + \omega^{mn} + \sum _{\gamma'}q_{\gamma'}\partial_{\gamma'} \varepsilon_{\bm k}^m+  \frac{1}{2}\sum_{\alpha', \beta'}q_{\alpha'} q_{\beta'}\,\partial_{\alpha'\beta'}\varepsilon^m_{\bm k} + i\eta}\\ \nonumber
&\times\Big[ 1- \sum_{\gamma''} \big\{mn \sin2\delta\theta_{\bm k}^{mn}  q_{\gamma''} \partial_{\gamma''}\theta_{\bm k}^m \big(1+(q_{\gamma''} \partial_{\gamma''} \theta_{\bm k}^m)^2 + \cdots\big)\\
&  +mn\cos 2\delta\theta_{\bm k}^{mn} \big( 1 - \{2  q_{\gamma''} \partial_{\gamma''} \theta_{\bm k}^m\}^2+\cdots\big)\big\}\Big].
\end{align}

Here we define $f^{mn} = f(\varepsilon^m_{\bm k}) - f(\varepsilon^n_{\bm k})$, $\partial_\gamma = \partial /\partial {k}_\gamma$ and $\delta\theta_{\bm k}^{mn} = \theta_{\bm k}^m - \theta_{\bm k}^n$, where $\theta_{\bm k}^n$ is the polar angle. % that measures the orientation of the Hamiltonian's pseudospin vector relative to the in-plane direction
Now, for a general two-band model, one can further simplify the Eq.~(\ref{eqn:polarizability_sim}) and it reduces to the form 

\begin{align} \nonumber\label{eqn:polarizability1_twoband} 
\text{P}^{mn}(\mathbf{q}\rightarrow 0, \omega) &=\frac{1}{2} \sum_{\bm k} 
\frac{f^{mn}+ \sum_\gamma q_\gamma \partial_{\gamma} f(\varepsilon_{\bm k}^m)}{\omega + \omega^{mn} + \sum _{\gamma'}q_{\gamma'}\partial_{\gamma'} \varepsilon_{\bm k}^m + i\eta}
\\ & \times \bigg[ 1- \sum_{\gamma''}mn ( 1 - \{2 q_{\gamma''} \partial_{\gamma''} \theta_{\bm k}^m\}^2+\cdots)\bigg],
\end{align}

where we consider $\theta^m_{\bm k}=\theta^n_{\bm k}$, and retain only leading-order terms in $\bm q$. %, as higher order contributions are neglected due to their negligible contributions (see Appendix~\ref{Appendix:A1}). %
Now the above equation can be decomposed into two parts: (i) intraband part and (ii) interband part. The intraband contribution to the polarizability, arising solely from the single-band dynamics, can be written by setting $m=n$ as 
\begin{align} \label{eqn:polarizability_low temperature}
   \text{P}^{mm}(\mathbf{q}\rightarrow 0,{\omega})
= - \sum_{\bm k}\sum_{\gamma,\gamma'}  \frac{{q}_\gamma {q}_{\gamma'}}{{\omega}^2} \frac{\partial {\varepsilon}^m_{\bm k}}{\partial {k}_\gamma}\frac{\partial {\varepsilon}^m_{\bm k}}{\partial {k}_{\gamma'}} \frac{\partial f({\varepsilon}^m_{\bm k})}{\partial {\varepsilon}^m_{\bm k}}.
\end{align}
Here,  we expand the factor $[\omega+\frac{\partial \varepsilon^m_{\bm k}}{\partial k _{\gamma'}} q_{\gamma'}]^{-1}= 1/\omega(1-\frac{\partial \varepsilon^m_{\bm k}}{\partial k _{\gamma'}}\frac{q_{\gamma'}}{\omega})$ and keep the non-zero leading order term in $\bm q$. It is clear from the above expression that the intraband part is a Fermi surface quantity due to the presence of $\frac{\partial f(\varepsilon^m_{\bm k})}{\partial \varepsilon^m_{\bm k}}$ and becomes $\mu$-dependent. 

Following Eq.~\eqref{eqn:polarizability1_twoband}, the interband part ($m \neq n$) of the polarizability $\text{P}^{mn}$, arising from the dynamics between two bands, can be expressed as 
\begin{align} 
\label{eqn:P+-_2a}\nonumber
 &\text{P}^{mn}(\bm q\rightarrow 0, \omega) =  2 \sum_{\bm k}\sum_{\gamma''}\big(  q_{\gamma''} \partial_{\gamma''} \theta_{\bm k}^m\big)^2\\\nonumber & \times~\frac{-1 +\Theta(\mu-\varepsilon_{\bm k}^{m})- \sum_\gamma q_\gamma \frac{\partial \varepsilon_{\bm k}^{m}}{\partial \bm k_\gamma}\delta(\mu -\varepsilon_{\bm k}^{m})}{\omega + \omega^{mn} + i\eta} \\ &=\text{P}^{mn}_{1} + \text{P}^{mn}_{2},
\end{align}
where, at low temperature regime, we consider doped Dirac-type systems in which the conduction band is electron-like and the valence band is fully filled (hole-like), i.e., $f(\varepsilon_{\bm k}^m) = \Theta(\mu-\varepsilon_{\bm k}^{m})$, $f(\varepsilon_{\bm k}^n) = 1$. 
Here, $\text{P}^{mn}_{1} = \sum\limits_{\bm k,\gamma''} \mathcal{A}_{\bm k, \gamma''}\big( \Theta(\mu-\varepsilon_{\bm k}^{m})-1\big)$ and $ \text{P}^{mn}_{2}=- \sum\limits_{ \bm k,\gamma, \gamma''}\mathcal{A}_{\bm k, \gamma''}  q_\gamma \frac{\partial \varepsilon_{\bm k}^{m}}{\partial \bm k_\gamma}\delta(\mu -\varepsilon_{\bm k}^{m})$ represent the chemical potential independent and dependent parts of the polarizability, respectively with $\mathcal{A}_{\bm k, \gamma''}=  \frac{2( q_{\gamma''} \partial_{\gamma''} \theta_{\bm k}^m)^2}{\omega + \omega^{mn} + i\eta}$.

We now discuss two limits arising from the competition between the chemical potential and band energy. In the first case $\mu>\varepsilon_{\bm k}^m$, $\Theta(\mu- \varepsilon_{\bm k}^m)=1$, therefore $\text{P}_1^{mn}$ term vanishes and the only interband contribution arises from the $\text{P}_2^{mn}$. On the other hand, for $\mu<\varepsilon_{\bm k}^m$, the Heaviside theta function becomes $\Theta(\mu - \varepsilon_{\bm k}^m)=0$ and consequently, both $\text{P}_1^{mn}$ and $\text{P}_2^{mn}$ contribute to the polarizability. Further, it is important to note that the chemical potential dependent part gives a non zero contribution in the case when $\mu=\varepsilon_{\bm k}^m$, due to the presence of the delta function. The chemical potential dependent $\text{P}^{mn}_{2}$ has not been introduced earlier and is one of the central results of our study.\\

\section{Polarizability of Dirac nodal line semimetal}
\label{Sec:III}
\subsection{Model Hamiltonian}

To compute the polarizability in 3D Dirac nodal line semimetals (DNLSMs), we consider an effective $\bm {k.p}$ two-band model Hamiltonian $ \mathcal{H}({\bm k}) =  \varepsilon_0\big[\space\big(\mathcal{\tilde{K}}-1\big) \space\sigma_x +  \space \gamma \tilde{k}_z\space\sigma_y + \tilde{M} \space \sigma_z\big]$. The third term of the Hamiltonian breaks the $\mathcal{PT}$ symmetry~\cite{Barati_prb2017, Wang_prb2021, Flores-Calderón_EL2023, Pandey_prb2024, Pandey_pssrrl2025}. The corresponding energy dispersion is 
\begin{eqnarray}
   & \tilde{\varepsilon}_{\bm k}^{\pm} =  \pm \sqrt{(\tilde{\mathcal{K}} - 1)^2+(\gamma \tilde{k}_z)^2 + \tilde{M}^2}.
\end{eqnarray}
Here, $ \tilde{\varepsilon}_{\bm k}^{\pm} = \varepsilon_{\bm k}^{\pm}/\varepsilon_0$, where $\varepsilon_0= k_0^2/(2m_e)$ is the energy associated with the nodal ring, $k_0$ is the nodal ring radius, and $m_e$ is the electronic mass. $\pm$ represents the conduction band and valence band respectively. Further, $\tilde{\mathcal{K}} =\mathcal{K}/k_0$ with
%$\mathcal{K}=\sqrt{k_x^2+k_y^2}$
$\mathcal{K}$ be the magnitude of the planar wave vector, $\tilde{k}_z=k_z/k_0$ and $\gamma=2 m_e v_z/ k_0$, where $v_z$ is the component of Fermi velocity along $z$-direction, $\sigma_{i}  (i \equiv x,y,z)$ refers to the Pauli matrices in the pseudospin basis. $\tilde{M}=M/\varepsilon_0$, having $M$ as the mass (gap) term, emerges experimentally from the application of an external magnetic field, pressure, stress, inversion-breaking uniaxial strain, etc.~\cite{Wang_prb2021, Flores-Calderón_EL2023, chiba_prb2017, Kot_prb2020, chen_prb2018, Rendy_jap2021, du_nrp2021}. For $\tilde{M}=0$, the nodal ring preserves all symmetry, such as $\mathcal{P}$, $\mathcal{T}$ and $\mathcal{PT}$.
%
%In addition, the eigenvector is $|u_{{\bm k}}^{n} \rangle = \left\{ \zeta^{-}_{\bm k} e^{-i\theta^n_{\tilde{\bm{k}}}}, ~\zeta^{+}_{\bm k}  \right\}^{T}$, 
    %
%where $\zeta_{\bm k}^{\pm}=\pm \frac{1}{\sqrt{2}}[1\pm \tilde{M}/\tilde{\varepsilon}_{\bm k}^{n}]^{1/2}$ and
For this model system, we find that the eigenfunctions yield the relation $\theta^{\pm}_{\tilde{\bm{k}}} = \tan^{-1} [\gamma \tilde{k}_z/ (\tilde{\mathcal{K}}-1)]$. This relation subsequently allows us to calculate the polarizability of the system using Eqs.~(\ref{eqn:polarizability_low temperature})-(\ref{eqn:P+-_2a}). %{\color{blue}Since the eigenvectors arise from the pseudospin structure for the model, it is same for $n = +$ and $n=-$ bands. Hence, although written with band index, in our two-band model $\theta_{\bm k}^{+} = \theta_{\bm k}^{-}$.}

We would like to point out that in this work, we focus on the regime defined by $\text{max}\{\sqrt{1-\frac{\tilde{M}}{\tilde{\mu}}}(\tilde{q}_\rho, \gamma \tilde{q}_z)\}<\tilde{\omega}<2\tilde{\mu}$, where $\tilde{q}_\rho = \sqrt{\tilde{q}_x^2 + \tilde{q}_y^2}$. In this parameter range, the imaginary part of the polarizability vanishes, while the real part remains finite. In the following subsections, we evaluate the real part of the polarizability arising from both interband and intraband processes for the modeled Dirac nodal-line semimetal, as well as for the representative materials Ca$_3$P$_2$ and ZrSiS.

\subsection{Interband Polarizability}\label{subsec:interband polarizability}
Now we will investigate the total interband polarizability which contains two parts: (i) $\text{P}^{mn}_2$ and (ii) $\text{P}^{mn}_1$.

\subsubsection{Chemical potential dependent contribution ($\text{P}^{mn}_2$)}

 The chemical potential dependent contribution $\text{P}_2^{mn}$ for considered DNLSM system consists of two parts: $\text{P}_2^{+-}$ and $\text{P}_2^{-+}$. The first part can be evaluated using Eq.~(\ref{eqn:P+-_2a}) as

\begin{align} \label{eqn:P_2^{+-}}
   \text{P}^{+-}_2=\int_{0}^{\infty} \frac{\tilde{\mathcal{K}}d\tilde{\mathcal{K}}}{4\pi^2}
 \frac{\tilde{q}_z \big[\gamma \tilde{q}_\rho\tilde{k}_{z0}-\gamma (\tilde{\mathcal{K}}-1) \tilde{q}_z\big]^2 }{(\tilde{\omega} + 2\tilde{\mu}+i\eta)(\tilde{\mu}^2 -\tilde{M}^2)^2} \Theta(\tilde{k}_{z0}^2),
\end{align}
where $\tilde{\mu}= \mu/\varepsilon_0$. To obtain the above equation, we use the cylindrical coordinate system and treat terms in the long-wavelength limit which yields $\sum_{\gamma''} \tilde{q}_{\gamma''} \partial_{\gamma''} \theta_{\tilde{\bm k}}^+=\frac{2}{\tilde{\varepsilon}_{\bm k}^2-\tilde{M}^2} \times (\gamma \tilde{k}_z \tilde{q}_\rho -\gamma \tilde{q}_z (\tilde{\mathcal{K}}-1))$ and perform the integration over $\tilde{k}_z$ using the property of the delta function.
%(1-\cos{2\delta\theta^{+-}_{\bm k  \bm q}})=
Here we define $\tilde{k}_{z0}=\sqrt{\tilde{\mu}^2-(\tilde{\mathcal{K}}-1)^2-\tilde{M}^2}$, and insert a function $\Theta(\tilde{k}_{z0}^2)$ to ensure the existence of real solutions for $\tilde{k}_z$. 
\begin{figure*}[t]
    \centering
    \includegraphics[width=16 cm]{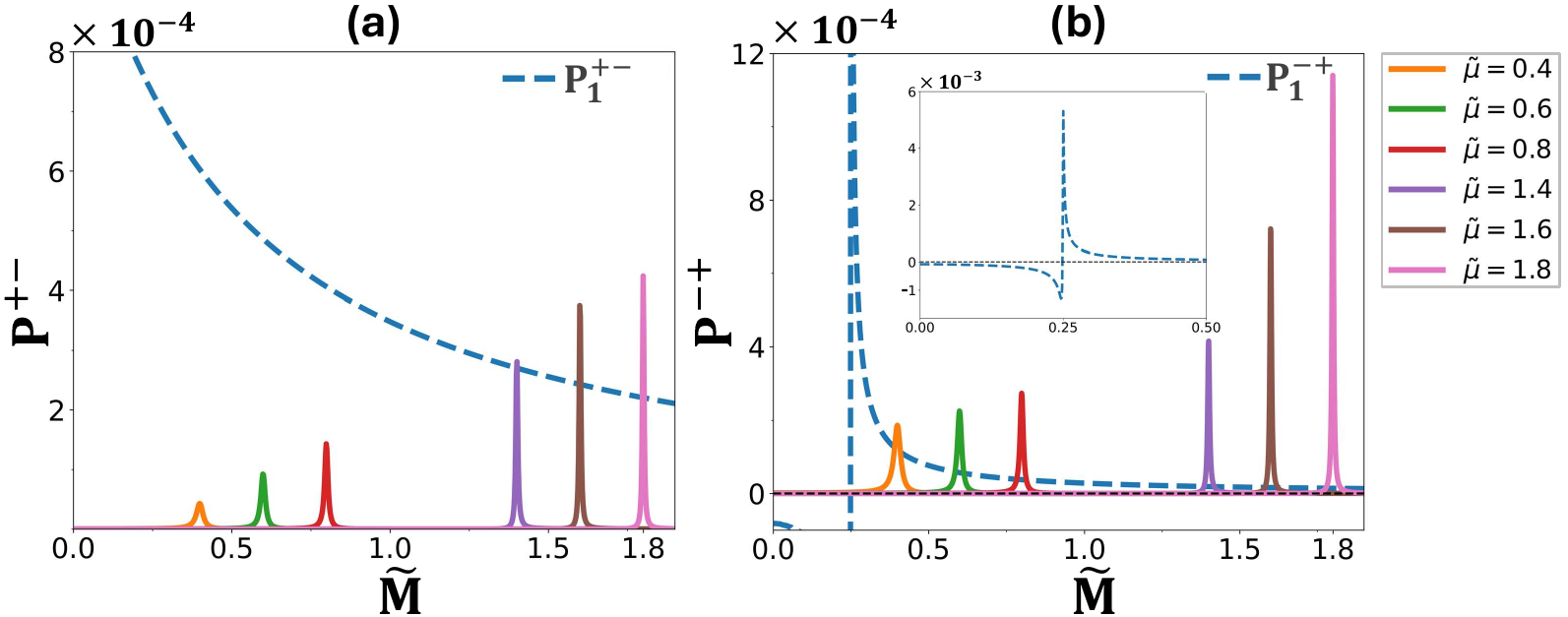}
    \caption{The plot illustrates the interband component of the polarizability (in units of $k_0^3/\varepsilon_0$) as a function of the mass parameter $\tilde{M}$ for different values of the chemical potential $\tilde{\mu}$. Panels (a) and (b) correspond to $\text{P}^{+-}$ and $\text{P}^{-+}$, respectively. The quantities $\text{P}_1^{+-}$ and $\text{P}_1^{-+}$ denote the $\tilde{\mu}$-independent parts of the interband polarizability in the DNLSM. The inset in panel (b) depicts the full variation of $\text{P}_1^{-+}$ with $\tilde{M}$ at a fixed frequency. The parameters used in the calculation are $\tilde{\omega} = 0.5$, $\tilde{q}_{\rho} = 0.01$, $\tilde{q}_{z} = 0.01$, and $\gamma = 2.8$.} 
    \label{fig:3}
\end{figure*}
It is noteworthy that $\text{P}^{+-}_2$, being proportional to $\tilde{q}_z$, vanishes in two-dimensional systems ($\tilde{q}_z \to 0$). Furthermore, it exhibits resonant behavior when the chemical potential approaches the bottom of the conduction band due to the presence of the $(\tilde{\mu}^2 - \tilde{M}^2)$ term. Since an exact evaluation of Eq.(~\ref{eqn:P_2^{+-}}) is cumbersome, we restrict our analysis to two limiting regimes: $\tilde{\mu}<\sqrt{1+\tilde{M}^2}$ and $\tilde{\mu}>\sqrt{1+\tilde{M}^2}$.\\
\textit{Case-I:}  $\tilde{\mu}<\sqrt{1+\tilde{M}^2}$, the integration limits for $\tilde{\mathcal{K}}$ lie between $-\tilde{\mu}$ to $\tilde{\mu}$. This yields 
\begin{align} \label{eqn:10}\nonumber
&\text{P}_2^{+-}=\frac{\tilde{q}_z}{24 \pi^2  (\tilde{\omega} + 2\tilde{\mu}+i\eta)(\tilde{\mu}^2 -\tilde{M}^2)^2}  \\ & \times\left[
\tilde{q}_\rho^2 \mathcal{B}_1(\tilde{M}, \tilde{\mu})+ \gamma \tilde{q}_z \left( \tilde{q}_\rho \mathcal{B}_2(\tilde{M}, \tilde{\mu}) + \gamma \tilde{q}_z \mathcal{B}_3(\tilde{M}, \tilde{\mu})\right)
\right].
\end{align}
Here, the detailed expressions of factors $\mathcal{B}_1(\tilde{M}, \tilde{\mu})$, $\mathcal{B}_2(\tilde{M}, \tilde{\mu})$ and  $\mathcal{B}_3(\tilde{M}, \tilde{\mu})$, are given in Appendix~\ref{Appendix:c}. Since $\mathcal{B}_1(\tilde{M}, \tilde{\mu})$, $\mathcal{B}_2(\tilde{M}, \tilde{\mu})$ and  $\mathcal{B}_3(\tilde{M}, \tilde{\mu})$ are linearly proportional to $\tilde{\mu}$, hence as the chemical potential $\tilde{\mu}\to 0$, the $\text{P}_2^{+-}$ vanishes. 
%Further, the part of the interband polarizability $\text{P}_2^{+-}$ shown in Eq.~\eqref{eqn:10} 
Further, as evident from the Eq.~\eqref{eqn:10}, it follows a different variation with chemical potential and mass term. %in comparison to the zero chemical potential case for the regime $\tilde{M}<1$, $\tilde{\mu}<1$. 
Specifically, the $\text{P}_2^{+-}$ scales as $\tilde{\mu}^{-1}$ when the Fermi level lies within the conduction band ($\tilde{\mu}>\tilde{M}$).
\\
\textit{Case-II:} $ \tilde{\mu}>\sqrt{1+\tilde{M}^2}$, the integration limits for $\tilde{\mathcal{K}}$ go from $-\sqrt{1+\tilde{M}^2}$ to $\tilde{\mu}$ and the solution turns out to be quite complicated. Thus, it is better to express it in the following form
\begin{align}\label{eqn:P_2^(+-)_case2}\nonumber
   &\text{P}_2^{+-} = \frac{\tilde{q}_z}{48 \pi^2 (\tilde{\omega} + 2\tilde{\mu}+i\eta)(\tilde{\mu}^2 -\tilde{M}^2)^2}\\&\times\left[
\tilde{q}_\rho^2 \mathcal{C}_1(\tilde{M}, \tilde{\mu})+ \gamma \tilde{q}_z \left( \tilde{q}_\rho \mathcal{C}_2(\tilde{M}, \tilde{\mu}) + \gamma \tilde{q}_z \mathcal{C}_3(\tilde{M}, \tilde{\mu})\right)
\right].
\end{align}
The explicit forms of $\mathcal{C}_1(\tilde{M},\tilde{\mu})$,  $\mathcal{C}_2(\tilde{M},\tilde{\mu})$ and  $\mathcal{C}_3(\tilde{M},\tilde{\mu})$ are quite large, and detailed expressions of these terms are provided in~\ref{Appendix:c}. 
For the chemical potential larger than the gap value and keeping the gap value smaller than unity,  $\text{P}_2^{+-}$ shows a dependence on $\tilde{\mu}^{-1}$.
   
It is important to note that the second part of the interband polarizability $\text{P}^{-+}_2$ can be obtained in similar way as $\text{P}^{+-}_2$. However, it yields an expression analogous to $\text{P}^{+-}_2$, with a slight change in the factor $\omega -2\tilde{\mu} + i\eta$ which is present in the denominator, thus gives slight variation in results as shown in the Fig.~\ref{fig:3}.% This factor gives the resonance at frequency equal to the gap between bands i.e, $\tilde{\omega} = 2\tilde{M}$. However, the other features remain qualitatively the same.

\subsubsection{Chemical potential independent contribution ($\text{P}^{mn}_1$)}
The conventional chemical potential independent part for DNLSMs arises from Eq.~(\ref{eqn:P+-_2a}),   

\begin{align}
  \text{P}_1^{+-} = -\int_{0}^{\infty} \int_{-\infty}^{+\infty}\frac{\tilde{\mathcal{K}}\,d\tilde{\mathcal{K}}\, d\tilde{k}_z}{4\pi^2}
 \frac{\big[\gamma \tilde{q}_\rho\tilde{k}_{z}-\gamma (\tilde{\mathcal{K}}-1) \tilde{q}_z\big]^2}{(\tilde{\omega} + 2\tilde{\varepsilon}_{\bm k}^{+}+i\eta)\big[(\tilde{\varepsilon}^{+}_{\bm k})^2 -\tilde{M}^2\big]^2}.    
\end{align}
It is to be noted that the above equation for $\text{P}_1^{+-}$ survives only for the case of $\Theta(\tilde{\mu} - \tilde{\varepsilon}_{\bm k}^+)=0$. 
While in the opposite case this contribution cancels with the contribution stemming from the term associated with Heaviside theta function. Further, the absence of Heaviside theta function term causes this part of the interband polarizability to become independent of $\tilde{\mu}$.  For the case of $\text{P}_1^{-+}$, the factor $\tilde{\omega} -2\tilde{\varepsilon}^{+}_{\bm k} + i\eta$ present in the denominator, gives a resonance at a frequency equal to the bandgap i.e, $\tilde{\omega} = 2\tilde{M}$.

In Fig.~\ref{fig:3}(a), we present the interband part of the polarizability $\text{P}^{+-}$ for DNLSM having two parts: $\text{P}_1^{+-}$ the chemical potential independent part shown by the dashed line and $\text{P}_2^{+-}$ the chemical potential dependent part displayed by the solid line. The $\tilde{\mu}$ independent part originates from transitions far away from the Fermi level and explains the kind of background interband response of the Dirac nodal line system. This contribution decreases with increasing $\tilde{M}$, exhibiting a $\tilde{M}^{-1}$ dependence, and scales with frequency as $\tilde{\omega}^{-1}$. Therefore, this contribution is significant at small band gap values and low frequencies. 
In addition, the $\tilde{\mu}$ dependent part shows resonant interband transitions when the chemical potential approaches the gap i.e., $\tilde{\mu}=\tilde{M}$, and the strength of the peaks increases with increasing $\tilde{\mu}$. 
However, for the chemical potential deeper in the conduction band ($\tilde{\mu}>\tilde{M}$), most of the interband transitions are Pauli blocked, thus reducing the response. It is important to note that the $\tilde{\mu}$ dependent resonant interband contribution stems from states near the band edges, where Pauli blocking is lifted, leading to the resonant response observed in this regime.

On the other hand, Fig.~\ref{fig:3}(b) illustrates the $\text{P}^{-+}$ part of the interband polarizability in a DNLSM. Here, the $\tilde{\mu}$ independent part of the interband polarizability exhibits resonance features followed by prominent transition peaks at $\tilde{M}=\tilde{\omega}/2$, indicative of Kramers-Kronig-type dispersive features. This is a universal feature of interband polarizability at the band edge.
 However, the peak at $\tilde{\omega}=2\tilde{M}$ is absent in the $\tilde{\mu}$ dependent component of the interband polarizability. The other peaks at $\tilde{\mu}=\tilde{M}$ follows a similar behavior as observed in Fig.~\ref{fig:3}(a), with slight differences arising due to the weighted factors. 
At a large value of $\tilde{M}$, the band gap becomes significantly larger, making the fixed photon energy $\tilde{\omega}$, insufficient to excite the electrons across the band gap. As a result, the interband transition becomes blocked, and the polarizability gradually diminishes.\\

\subsection{Intraband Polarizability} \label{subsec:Intraband Polarizability}
The intraband polarizability comprises two components: $\text{P}^{++}$ comes purely from the conduction band, and $\text{P}^{--}$ from the valence band.
%
%
%First, the $\text{P}^{++}$ intraband part of the polarizability coming from Eq.~\eqref{eqn:polarizability_low temperature} by setting $m=n=+$.
%
It is clear from Eq.~\eqref{eqn:polarizability_low temperature}, the intraband part of the polarizability shows a quadratic dependence on $|\tilde{\bm q}|$ and inverse square law with $\tilde{\omega}$.
Further on approximating the energy derivative of the Fermi Dirac distribution function $-\delta(\tilde{\mu} - \tilde{\varepsilon}_{\bm k})$ at the low-temperature limit and performing summations over $\gamma, \gamma^\prime = \{x,y,z\}$, the intraband polarizability arising from the conduction band for the modeled DNLSM simplifies to
 \begin{align}\label{eqn:P++_main}
  &   \text{P}^{++} = \frac{1}{\tilde{\omega}^2} \int _{0}^{\infty} \frac{d\tilde{\mathcal{K}} \tilde{\mathcal{K}}}{4 \pi^2\gamma^2}\frac{1}{\tilde{\mu}\tilde{k}_{z0}}
     \big[(\mathcal{\tilde{K}}-1)^2 \tilde{q}_{\rho}^2+ \gamma^2 \tilde{k}_{z0}^2 \tilde{q}_z^2\big] \Theta(\tilde{k}_{z0}^2).
 \end{align}
Here we use the properties of the delta function to perform the integration over $\tilde{k}_z$ and introduce $\Theta(\tilde{k}_{z0}^2)$ which guarantees the existence of the real solutions for $\tilde{k}_z$. 
However, analogous to the interband part of the polarizability, here we have two limiting cases, with $\tilde{\mu}<\sqrt{1+\tilde{M}^2}$ or $\tilde{\mu}>\sqrt{1+\tilde{M}^2}$.\\ % 
\textit{Case-I:} $\tilde{\mu}<\sqrt{1+\tilde{M}^2}$, $\text{P}^{++}$ becomes
\begin{align}
\label{eqn:P++L}
    &\text{P}^{++}= \frac{1}{4 \pi^2 \gamma^2 \tilde{\mu} \tilde{\omega}^2}[\mathcal{F}(\tilde{M}, \tilde{\mu}) (\tilde{ q}_{\rho}^2 + \gamma^2 \tilde{q}_z^2) + 2 \tilde{M} \tilde{\mu} \gamma^2 \tilde{q}_z^2],
\end{align}
where $\mathcal{F}(\tilde{M},\tilde{\mu})= - \big[(\tilde{M}^2 -\tilde{\mu}^2)~\text{tan}^{-1}{(\tilde{\mu}/\tilde{M})} + \tilde{M}\tilde{\mu} \big]$. 
%
%To understand the behavior of the intraband polarizability given in Eq.~\eqref{eqn:P++L}, we can categorize the analysis into two different regimes depending on the variation of $\tilde{M}$ and $\tilde{\mu}$. First, we consider the regime $\tilde{M}<1$, $\tilde{\mu}<1$.
Here, the intraband polarizability follows $\tilde{\mu}~\text{tan}^{-1}(\tilde{\mu}/\tilde{M})$ behavior when $\tilde{\mu}>\tilde{M}$ as depicted in Table~\ref{tab:SP}.\\ %while for $\tilde{\mu}<\tilde{M}$ it yields linear dependence in $\tilde{M}$.
%
%In the second regime, defined by $\tilde{M}>1$, $\tilde{\mu}<\tilde{M}$, two scenarios arise $\tilde{\mu}<1$ and $\tilde{\mu}>1$, which exhibits a linear dependence on $\tilde{M}$.\\ 
%
\textit{Case-II:} $\tilde{\mu}>\sqrt{1+\tilde{M}^2}$, the intraband polarizability component $\text{P}^{++}$ becomes,
\begin{align} \label{P++_1}
  &  \text{P}^{++}= \frac{1}{24 \pi^2 \gamma^2 \tilde{\mu} \tilde{\omega}^2}\bigg[\mathcal{A}_1(\tilde{M}, \tilde{\mu})~\tilde{ q}_{\rho}^2 +  \mathcal{A}_2(\tilde{M}, \tilde{\mu})~\gamma^2 \tilde{q}_{z}^2\bigg].
\end{align}
Here, the explicit forms of $\mathcal{A}_1(\tilde{M},\tilde{\mu})$, and $\mathcal{A}_2(\tilde{M},\tilde{\mu})$ are quite cumbersome and are provided in the Appendix~\ref{Appendix:A}. However, we discuss these factors within the limiting regime. 
In the regime %$\tilde{M}<1$, $\tilde{\mu}>1$ which ensures the chemical potential is greater than the mass value, i.e., 
$\tilde{\mu}>\tilde{M}$, both factors $\mathcal{A}_1(\tilde{M},\tilde{\mu})$ and $\mathcal{A}_2(\tilde{M},\tilde{\mu})$ show dependencies on $\tilde{\mu}^2 \tan^{-1} (\tilde{\mu}/\tilde{M})$ and $\tilde{\mu}^3$. Therefore, the intraband polarizability within this regime follows a pattern depending on the competition between $\tilde{\mu} \tan^{-1} (\tilde{\mu}/\tilde{M})$ and $\tilde{\mu}^2$ and shown in Table~\ref{tab:SP}.
\begin{table*}[t]
    \centering
    \fontsize{9pt}{9pt}\selectfont  % Compact font
    \renewcommand{\arraystretch}{1.8}  % Adjust vertical spacing
    \caption{The table shows the power-law dependence and the location of observed peak for the intraband, $\tilde{\mu}$ dependent and independent part of interband polarizability depending on quantities such as mass ($\tilde{M}$), chemical potential ($\tilde{\mu}$) in distinct regimes of interest, respectively. Where $*$ stands for the competition between two scaling factors.}
    \label{tab:SP}
    \begin{ruledtabular}
    \begin{center}
    \begin{tabular}{ccc|ccc}
        \multirow{2}{*}{\makecell{Limiting Cases}} 
        & \multicolumn{2}{c|}{\makecell{Intraband\\Polarizability}}
        & \multicolumn{3}{c}{\makecell{Interband Polarizability}} \\ 
        \cmidrule(lr){2-3} \cmidrule(lr){4-6}
        & \makecell{$\tilde{\mu}$-dependence} 
        & \makecell{$\tilde{\omega}$-dependence} 
        & \makecell{$\tilde{\mu}$-dependent\\part} 
        & \makecell{$\tilde{\mu}$-independent\\part} 
        & \makecell{$\tilde{\omega}$-dependence} \\
        \hline
        $\tilde{\mu}<\sqrt{1+\tilde{M}^2}$ $(\tilde{\mu}>\tilde{M})$ 
        & $\tilde{\mu}\tan^{-1}(\tilde{\mu}/\tilde{M})$ & $\tilde{\omega}^{-2}$ 
        & $\tilde{\mu}^{-1}$ & -- & $\tilde{\omega}^{-1}$ \\

        $\tilde{\mu}>\sqrt{1+\tilde{M}^2}$ $(\tilde{\mu}>\tilde{M})$ 
        & \makecell{$\tilde{\mu}\tan^{-1}(\tilde{\mu}/\tilde{M})$ \\ \& $\tilde{\mu}^2{^*}$} 
        & $\tilde{\omega}^{-2}$ 
        & $\tilde{\mu}^{-1}$ & -- & $\tilde{\omega}^{-1}$ \\

        $\tilde{\mu}=\tilde{M}$ 
        & no peak & $\tilde{\omega}^{-2}$ 
        & resonant peak & no peak & $\tilde{\omega}^{-1}$ \\
    \end{tabular}
    \end{center}
    \end{ruledtabular}
\end{table*}
\begin{figure}[t]
    \centering
    \includegraphics[width=7.5 cm]{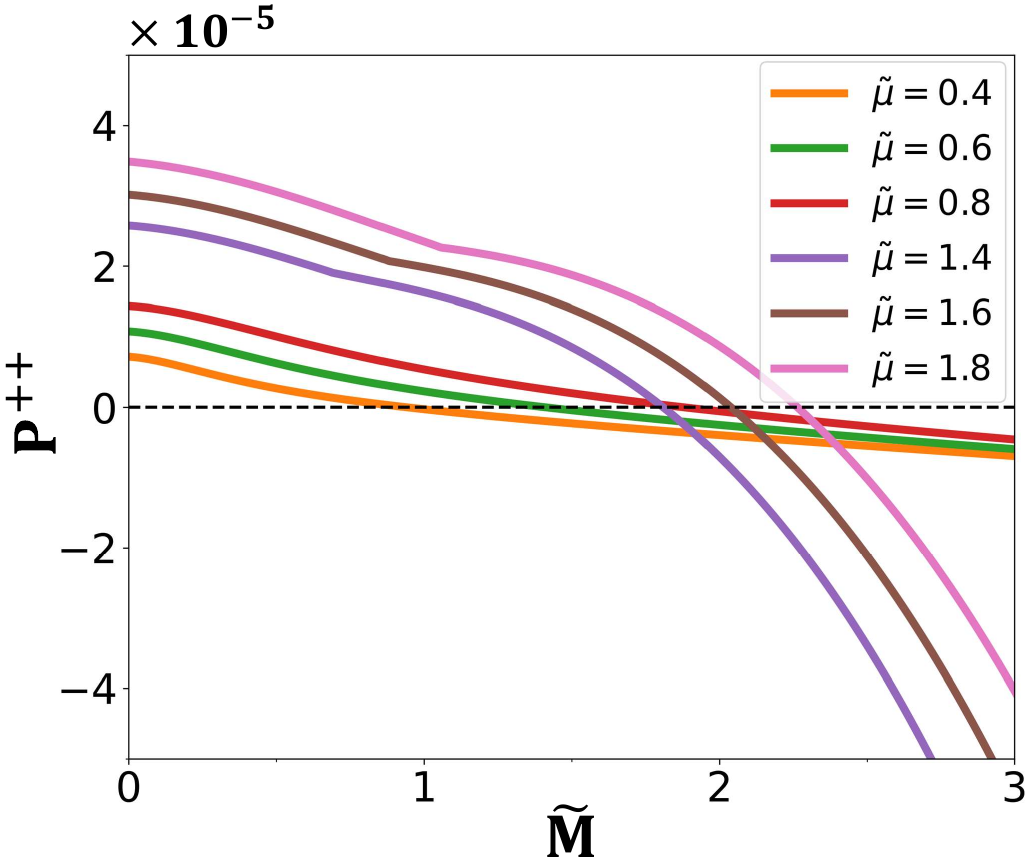}
    \caption{The $\text{P}^{++}$ component of the intraband polarizability (in units of $k_0^3/\varepsilon_0$) as a function of mass $\tilde{M}$ for different values of $\tilde{\mu}$ is depicted. Here, we fix $\tilde{\omega}=0.5$, $\tilde{q}_{\rho} =0.01$, $\tilde{q}_{z} = 0.01$ and $\gamma = 2.8 $.} 
    \label{fig:1}
\end{figure}

In Fig.~\ref{fig:1}, we demonstrate the variation of the intraband part of the polarizability, $\text{P}^{++}$ with the mass $\tilde{M}$ for different values of the chemical potential $\tilde{\mu}$. Here, we observe two different patterns depending on the chemical potential taken into account, as previously discussed in Eqs.~\eqref{eqn:P++L} and~\eqref{P++_1}. 
%
%In the regime $\tilde{\mu}<1$, 
At low $\tilde{\mu}$, the chemical potential lies near the band edge, resulting in a reduced intraband contribution that is highly sensitive to the mass parameter $\tilde{M}$. As $\tilde{\mu}$ increases and moves deeper into the conduction band, the density of states at the chemical potential increases, enhancing the intraband response and modifying its dependence on $\tilde{M}$. Moreover, the analytical form of $\text{P}^{++}$ changes between the regimes $\tilde{\mu}<\sqrt{1+\tilde{M}^2}$ and $\tilde{\mu}>\sqrt{1+\tilde{M}^2}$, leading to different power-law behaviors summarized in Table~\ref{tab:SP}. This explains the change in the trend of $\text{P}^{++}$ with varying $\tilde{\mu}$, as observed in Fig.~\ref{fig:1}.
Further, the intraband polarizability yields a finite value in the absence of the mass term i.e., at $\tilde{M}=0$. This behavior arises due to the finite density of states around the protected nodal ring and indicates metallic nature of the system. 
On turning the mass term on, $\tilde{M}\neq0$, the intraband polarizability decreases, followed by a sign change and linear variation with $\tilde{M}$ in case $\tilde{\mu}<1$. However, for $\tilde{\mu}>1$ case, beyond this point, the system follows a different power law on $\tilde{M}$ due to the interplay between the contributions associated with perpendicular and parallel $\tilde{\bm q}$ components. The resulting behavior is controlled by the position of the chemical potential (Fermi level) within the conduction band and in the gaped region. Here, the % location of the phase change where the intraband polarizability vanishes occurs at $\tilde{M}$ higher than $\tilde{\mu}$. This 
sign-flipping behavior indicates the transition from the metallic to the insulating region. At $\tilde{M}=\tilde{\mu}$, the Fermi level touches the edge of the conduction band. In this regime, intraband polarizability $\text{P}^{++}$ does not exhibit any resonance feature. This follows directly from Eq.~\eqref{eqn:P++_main}, where the intraband response depends smoothly on the derivative of the Fermi Dirac distribution and scales as $\tilde{\omega}^{-2}$, without having any singular factor.%   
%It is important to emphasize that in the strict insulating regime $-M<\mu<M$, the conduction band remains completely empty while the valence band is fully occupied. Consequently, $\partial f/\partial \varepsilon=0$ and both $P^{++}$ and $P^{--}$ vanish identically, in agreement with Eq.~(2) and with previous studies on nodal-line systems~\cite{Yan_prb2016}. 
%{\color{blue}In our analytical expression for $\text{P}^{++}$ [Eq.~\eqref{eqn:P++_main}], the apparent sign change observed in Fig.~\ref{fig:1} does not imply a finite intraband response inside the gap. Instead, it originates from the analytic continuation of terms such as $\tan^{-1}(\tilde{\mu}/\tilde{M})$, which interpolate smoothly between the metallic regime ($\tilde{\mu}>\tilde{M}$) and the insulating regime ($\tilde{\mu}<\tilde{M}$). Physically, this behavior reflects the vanishing density of states at the band edge, with the intraband contribution continuously decaying to zero as $\tilde{\mu} \to\tilde{M}$. The sign reversal in Fig.~\ref{fig:1} should therefore be understood as a crossover feature rather than a physical negative intraband response.}
%
%
%
%Additionally, the magnitude of the intraband polarizability also increases with the chemical potential, while the qualitative behavior remains unchanged. 
 Moreover, an increase in $\tilde{\mu}$ enhances the overall magnitude of the response and shifts the location of both the sign change and dip to the higher $\tilde{M}$. Despite these shifts, the qualitative behavior of the system remains unchanged. These results are consistent with the power law dependence discussed in Table~\ref{tab:SP}.
Furthermore, the other component of the intraband polarizability $\text{P}^{--}$ yields zero results in the low-temperature limit due to the vanishing difference between the Fermi Dirac distribution functions. More details are provided in the Appendix~\ref{Appendix:B}.\\

%
%In the regime $\tilde{\mu}>\sqrt{1+\tilde{M}^2}$, there can be two possibilities: one when $\tilde{M}<1$, $\tilde{\mu}>1$, which ensures that the chemical potential is greater than the mass term $\tilde{\mu}>\tilde{M}$ (the Fermi surface is no longer a torus), as discussed in Eq.~\eqref{P++_1}. 
%
%Here, at $\tilde{M}=0$ the intraband polarizability remains non-zero due to the finite density of state at the Fermi energy (shows a metal-like screening).
%In addition, on moving towards higher mass value and from $\tilde{M} < 1$ to $\tilde{M} >1$ regime, the polarizability decreases, followed by a sign change.
% 
%

%
\begin{figure}[t]
    \centering
    \includegraphics[width=7.5 cm]{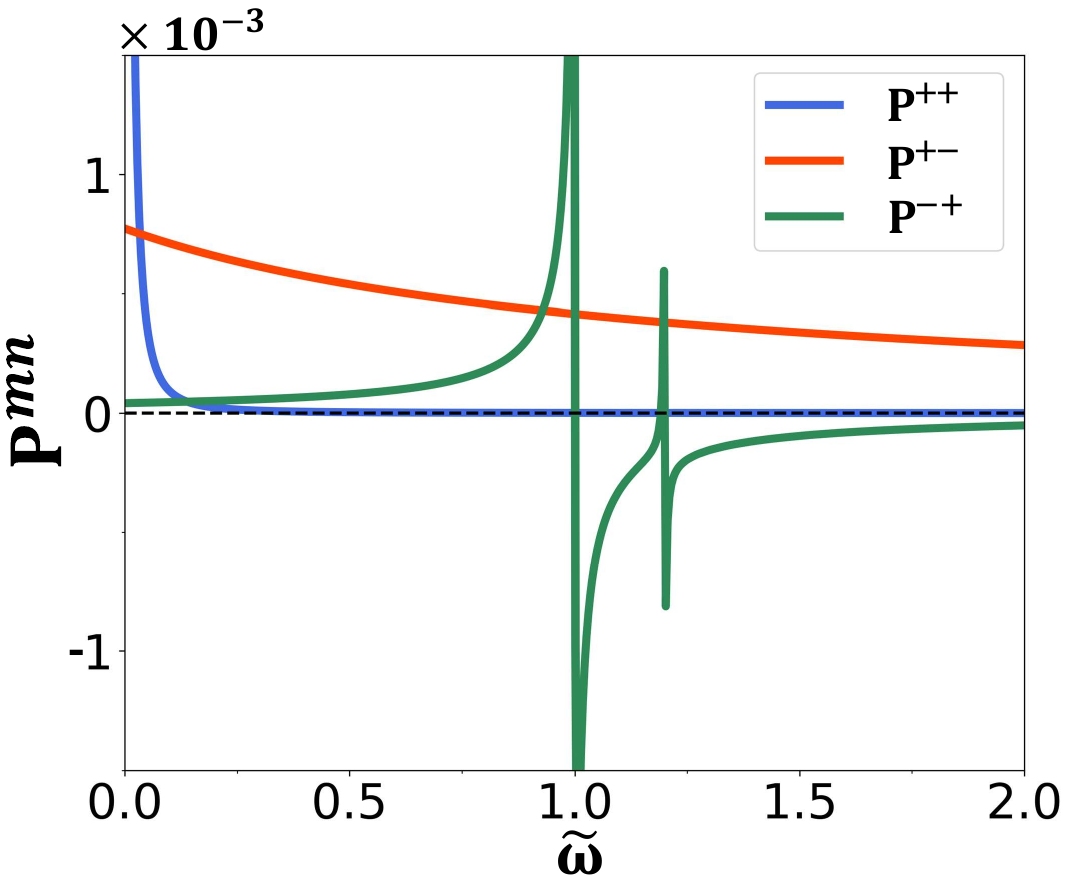}
    \caption{Plot depicts the comparison between the intraband and the interband components of the polarizability (in units of $k_0^3/\varepsilon_0$) with the frequency $\tilde{\omega}$ at fixed $\tilde{\mu} = 0.6$, $\tilde{M}=0.5$, $\tilde{q}_{\rho} =0.01$, $\tilde{q}_{z} = 0.01$ and $\gamma = 2.8 $.} 
    \label{fig:4}
\end{figure}

Fig.~\ref{fig:4} presents a comparison plot between the intraband ($\text{P}^{++}$) and the interband ($\text{P}^{+-}$ and $\text{P}^{-+}$) contributions to the polarizability of the DNLSM as a function of frequency $\tilde{\omega}$. We find the dominating nature of the intraband polarizability of the DNLSMs in a lower frequency regime. The resulting intraband feature shows the Drude-like nature and scales as $\tilde{\omega}^{-2}$. However, as the frequency increases into the intermediate and high frequency regimes, the interband contributions become increasingly significant and eventually dominate the overall polarizability of the DNLSM. In addition, a distinct transition peak emerges which originates from the interband component of the polarizability ($\text{P}_2^{-+}$), and it signifies the optical absorption, a similar resonance peak at $\tilde{\omega}=2\tilde{\mu}$ has been observed in Ref.~\cite{Barati_prb2017}. The power law dependence in chemical potential ($\tilde{\mu}$) and frequency ($\tilde{\omega}$) for the intraband and interband polarizability are summarized in Table~\ref{tab:SP}.
\section{Dielectric function} \label{Sec:dielectric function}

The dielectric function within the random phase approximation (RPA) is defined as~\cite{giuliani_cup2008, Hwang_prb2007, Rahimpoor_prb2024, Yan_prb2016,  Liu_prb2014}
\begin{align}\label{eqn:dielectric}
    \epsilon^{RPA}(\tilde{\bm q}, \tilde{\omega}) = 1- \sum_{m,n} v(\tilde{\bm q}) \text{P}^{mn}(\tilde{\bm q}, \tilde{\omega}),
\end{align}
where, $\sum\limits_{m,n}\text{P}^{mn}(\tilde{\bm q}, \tilde{\omega})= \text{P}^{mm}(\tilde{\bm q}, \tilde{\omega})\delta_{mn}+\sum\limits_{m\neq n}\text{P}^{mn}(\tilde{\bm q}, \tilde{\omega})$ with  $\text{P}^{mm}(\tilde{\bm q}, \tilde{\omega})= \zeta(\tilde{\bm q})/\tilde{\omega}^2$. Here, $\zeta(\tilde{\bm q})$ is the frequency independent part arising from Eq.~\eqref{eqn:P++_main} and $\text{P}^{mn}$ is the interband part of the polarizability. Furthermore the $v(\tilde{\bm q})= 4 \pi e^2/k_b k_0^2\tilde{\bm q}^2$, is the Coulomb potential in momentum space, where $k_b$ is the background dielectric constant and $\tilde{q}=\sqrt{\tilde{q}_\rho^2+\tilde{q}_z^2}$.
In the case of plasmon frequency,
\begin{align}
    0 = 1- v(\tilde{\bm q}) \big[\text{P}^{mm}(\tilde{\bm q}, \tilde{\omega})+\sum_{m\neq n}\text{P}^{mn}(\tilde{\bm q}, \tilde{\omega})\big].
\end{align}
It is important to note that the plasmon frequency is determined by the condition where the dielectric function becomes zero, i.e., $   \epsilon^{RPA}(\tilde{\bm q}, \tilde{\omega}) =0$ . This condition is stable in the region where the intraband contribution has a dominant effect, specifically in $\tilde{\omega}<\tilde{\mu}$ region. Hence, in this limit, the interband contribution to the polarizability becomes frequency independent and the plasmon frequency is given by the expression, 
\begin{align}\label{eqn:plasmonfrequency}
      \omega_p= \sqrt{\frac{v(\tilde{\bm q})\zeta(\tilde{\bm q})}{1- v(\tilde{\bm q}) \sum_{m\neq n}\text{P}^{mn}(\tilde{\bm q})}}. 
\end{align}
Here, the polarizability comprises contributions associated with the $\tilde{q}_\rho$ and $\tilde{q}_z$ components of the wave vector. To understand the plasmon frequency variation for DNLSMs stemming from the $\tilde{q}_\rho$ and $\tilde{q}_z$ components of the wave vector, we add the analysis below.  

In the limit $\tilde{q}_z\to0$, the plasmon frequency becomes
 \begin{align}
      \omega_{p,\rho}= \sqrt{\frac{v(\tilde{ q}_\rho)\zeta(\tilde{q}_\rho)}{1- v(\tilde{q}_\rho) \sum_{m\neq n}\text{P}_1^{mn}(\tilde{q}_\rho)}}. 
\end{align}

In this case, the contribution to the plasmon frequency comes from the intraband and the $\text{P}_1^{+-}$ part of the interband polarizability, as the $\text{P}_2^{+-}$ part is $\propto \tilde{q}_z$ (as shown in Eq~\eqref{eqn:P_2^{+-}}), thus vanishes in this limit. Furthermore, the Fermi energy dependence solely arises from the intraband component, whereas the $\text{P}_1^{+-}$ component remains independent of the chemical potential.

In contrast, in the case of the plasmon frequency coming from the $\tilde{q}_z$ component ($\tilde{q}_\rho\to0$), we obtain
\begin{align}
      \omega_{p, z}= \sqrt{\frac{v(\tilde{ q}_z)\zeta(\tilde{q}_z)}{1- v(\tilde{q}_z) \sum_{m\neq n}\text{P}^{mn}(\tilde{q}_z)}}. 
\end{align}
Here, the behavior of the plasmon frequency depends on both the variation of the intraband and interband parts of the polarizability in DNLSM, as discussed in the Sec.~\ref{subsec:Intraband Polarizability} and Sec.~\ref{subsec:interband polarizability}. 

The plasmon frequency, in the presence of both components of wave vector (i.e., $\tilde{q}_\rho$ and $\tilde{q}_z$), shows a complex dependency on the wave vector. However, the leading order contribution with an individual component $\tilde{q}_\rho$ and $\tilde{q}_z$ to the polarizability reveals the wave vector independent and dependent behavior of the plasmon frequency, respectively. 

\section{Experimental relevance and numerical estimation} \label{Sec:Experimental relevance}
Experimentally, the dielectric function of materials can be determined using techniques such as optical spectroscopy in the low-momentum regime and electron energy loss spectroscopy in high momentum regime. Since polarizability is a more microscopic quantity and is not directly measurable, it is typically calculated from the dielectric function. First, the intraband and interband contributions to the polarizability and to the dielectric function can be distinguished by modulating the applied oscillating field. 
Further, the separation of the resonant and non-resonant parts of the interband polarizability can be achieved by systematically tuning the parameters such as the chemical potential, frequency, and gap. These can be controlled using methods like electrostatic gating or doping (for chemical potential), optical excitation (for frequency), and strain engineering (for the gap). Resonance occurs as the Fermi level touches the bottom of the conduction band. However, on residing the Fermi level in the band gap, such a resonance feature does not occur. Therefore, by exploiting the correlation of the resonance peak with the Fermi level, one can easily distinguish the feature related to chemical potential dependent interband polarizability from other parts. In particular, the peaks led by interband transitions for polarizability at $\tilde{\omega} = 2\tilde{M}$ and $\tilde{\mu} = \tilde{M}$ can be probed by modulating the external applied field and doping. Here, the position and strength of such features are sensitive to the band gap, making it an important parameter for the polarizability.

For the numerical estimation of the polarizability of Dirac nodal line semimetals, we consider Ca$_3$P$_2$ and ZrSiS as promising materials due to their nodal line structure and tunable transport and optical properties.
For Ca$_3$P$_2$, we have used the DFT fitted parameters, such as radius of the nodal ring $k_0 \approx 0.206 \AA^{-1}$ and the energy associated with the nodal ring $\varepsilon_0 \approx 0.184$ eV and the parameter $\gamma \approx 2.80$~\cite{Xie_APLM2015, Chan_prb2016}. At a frequency $\tilde{\omega}=0.5$ (where $\tilde{\omega}=\omega/\varepsilon_0$), $\tilde{\mu} = 1.4$ and $\tilde{M} = 1.4$, the intraband polarizability is $\text{P}^{mm}= 0.05\times10^{-5}$ eV$^{-1}$$\AA^{-3}$, while the interband part $\text{P}^{mn} = \text{P}_1^{mn} + \text{P}_2^{mn} \approx 0.4 \times 10^{-4}$ eV$^{-1}$$\AA^{-3}$. Here, the $\tilde{\mu}$ independent and dependent parts, $\text{P}_1^{mn}$ and $\text{P}_2^{mn}$ respectively, relate as $\text{P}_2^{mn}\approx 5 \text{P}_1^{mn}$, and we find $\text{P}^{mn}\approx 80\text {P}^{mm}$. In this case, the dielectric function value is $| \epsilon^{RPA}(\bm q, \omega)|\approx10$.
In the case of ZrSiS, we consider $k_0 \approx 4.3 \AA^{-1}$, $\varepsilon_0 \approx 70$ eV and $\gamma \approx 0.20$~\cite{Schilling_prl2017}. Here, at frequency $\tilde{\omega}=0.5$, $\tilde{\mu} = 1.4$ and $\tilde{M} = 1.4$, the intraband part of polarizability yields $\text{P}^{mm}= 3\times10^{-4}$eV$^{-1}$ $\AA^{-3}$, while the interband part gives $\text{P}^{mn}= 1.9 \times 10^{-4}$eV$^{-1}$ $\AA^{-3}$. Here the $\text{P}_2^{mn}\approx 8 \text{P}_1^{mn}$ and the $\text{P}^{mm}\approx 1.5\text {P}^{mn}$. Further, the dielectric function becomes $| \epsilon^{RPA}(\bm q, \omega)|\approx 0.4$. It shows that, in the case of ZrSiS, the intraband and interband contributions are comparable to each other. However, the interband to intraband ratio is not large as compared to Ca$_3$P$_2$ due to large nodal ring radius. These estimates infer that such variations can be significantly observed as well with materials having a nodal ring radius smaller than that of ZrSiS.   

We would like to point out that at finite and large momentum transfer ($q\gg k_F$), the response becomes highly nonlocal and material specific, where the Taylor expansion of energy dispersion, the Fermi–Dirac distribution, and the overlap matrix elements are no longer valid.  
Although this regime creates rich effects such as Landau damping (decay into electron-hole excitations when energy-momentum conservation is satisfied) and single particle excitations, its description lacks universality: the results depend on band structure details, matrix element suppression, and local field corrections. As a consequence, short-wavelength polarizability does not provide a simple, generalizable picture, but rather requires case-by-case numerical treatment.

On the other hand, in the long wavelength ($\bm q\to0$) limit, the above expression can be simplified by using the Taylor series expansion of the respective terms. The long wavelength limit is physically relevant for many theoretical and experimental probes, such as optical conductivity, dielectric function, screening effect, and plasmon frequency (collective mode). Further in this limit, universal power laws and symmetry determine responses, e.g, Drude peak, interband transitions. In conclusion, the long wavelength limit makes the analysis simple and connects directly to the physical processes where momentum transfer is negligible, and this is the regime where polarizability and plasmonic features have the greatest physical significance.

In the long wavelength regime ($\bm q \to 0$) for nodal line semimetals, the condition $q\ll k_0$, where $k_0$ is the nodal ring radius, must be satisfied. To ensure the validity of this limit, we have considered $q_\rho = 0.01 k_0$ and $q_z = 0.01 k_0$, which firmly places the analysis within the long wavelength limit. Further, for the materials such as Ca$_3$P$_2$ and ZrSiS the nodal ring radius ($k_0$) values are $0.206 \AA^{-1}$ and $4.3 \AA^{-1}$ respectively. 
The random phase approximation (RPA) is appropriate for our analysis because NLSMs generally exhibit weak or moderate electron–electron interaction strength and a linear density of states, which suppress strong correlation effects near the nodal ring. In addition, the dielectric response studied here concerns the long-wavelength and low $q$ regime, where RPA is known to provide an accurate description of screening and collective modes. The plasmon energies in these materials typically lie in a range where vertex corrections and local-field effects are small, making RPA a reliable approximation for capturing both intraband and interband contributions.

At the same time, the RPA has some limitations. It neglects self-energy corrections and short-range correlation effects. When electron-electron interaction is strong, vertex corrections become large and one has to go beyond the RPA approximation. At large momentum transfer $q$, the induced charge density becomes nonuniform and the linear response assumption underlying RPA breaks down. This causes shifts in the plasmon energy dispersion, but qualitatively the results may not change. 

\section{Summary}\label{Sec:summary}
In summary, we have systematically revisited the polarizability of three-dimensional Dirac nodal line semimetals within the random phase approximation, uncovering several unconventional features that enrich the understanding of screening and collective excitations in these systems. A central outcome of our study is the identification of a novel interband contribution that acquires explicit chemical-potential dependence at low temperatures and exhibits cubic wave-vector scaling. This resonant term, which has no analogue in two-dimensional systems, could generate a distinctive optical fingerprint as the chemical potential approaches the band edge, thereby offering a powerful route for experimental detection and control.

Our analysis highlights the highly anisotropic nature of screening in 3D NLSMs. In particular, the emergence of a $q_z q^2$-dependent interband term underscores the role of out-of-plane momentum transfer, setting 3D NLSMs apart from their isotropic 2D counterparts. This anisotropic contribution can even surpass the conventional quadratic terms, leading to qualitatively new dielectric behavior. The frequency dependence of the polarizability further emphasizes the strong interplay between intraband and interband processes: intraband channels dominate at low frequencies, while interband transitions govern the intermediate and high-frequency regimes, reflecting strong coupling between bands near the nodal ring. We have also shown that introducing a $\mathcal{PT}$-symmetry–breaking mass term fundamentally alters the polarizability. Besides opening a gap and suppressing selected interband channels, the mass term induces a sign reversal in the intraband contribution, which can act as a marker of a metal–insulator transition. This establishes a direct and tunable connection between gap engineering and dielectric response, which can be probed experimentally via strain, pressure, or external fields.

To demonstrate the applicability of our theory, we have provided material-specific estimates for Ca$_3$P$_2$ and ZrSiS, both of which reveal strong tunability of the predicted effects. These case studies not only validate the general framework but also suggest promising candidates for experimental verification. Finally, one can extend the present framework to investigate the effect of tilt~\cite{Pandey_njp2025}, which breaks the inversion symmetry and can be further generalized for the analysis of other 3D systems~\cite{chang_arxiv2025, Chang_prb2017}.
The observed resonant feature in the chemical potential dependent part of interband polarizability can be used for dielectric engineering, anisotropic plasmonics and gated optoelectronic modulation~\cite{Xue_prl2021, Chorsi_afm2022, Chi_am2019, Zhou_aem2020, Shao_sa2022, Stauber_jpcm2014}.

\section{Acknowledgment}
This work is financially supported by Anusandhan National Research Foundation under project number SUR/2022/000289.

\onecolumngrid
\appendix
\section{Derivations of the intraband and interband polarizability expression} \label{Appendix:A1}
The general form of polarizability for the non-interacting system is
\begin{align} \label{eqn: polarizability}
    &\text{P}^{mn}(\bm q, \omega) = \frac{1}{2 v} \sum_{\bm k} \frac{f( \varepsilon^m_{\bm k+\bm q})-f(\varepsilon^n_{\bm k})}{\omega+ \varepsilon^m_{\bm k+\bm q}-\varepsilon^n_{\bm k}+i \eta} |\langle m, \bm k +\bm q|n, \bm k\rangle|^2.
\end{align}
Where, $|\langle m, \bm k +\bm q|n, {\bm k} \rangle|^2= 1+ mn \cos{2 \delta \theta_{\bm{kq}}^{mn}}$, where $\delta\theta_{\bm{kq}}^{mn}=\theta_{\bm k +\bm q}^m-\theta_{\bm k}^n$ is the angle between $\bm k$ and $\bm k + \bm q$ and lastly $\eta\rightarrow 0^+$ is a small positive infinitesimal number.   
In the long wavelength ($\bm q\to0$) regime, the terms such as energy dispersion, Fermi–Dirac distribution function, and form factors (overlap matrix elements) can be expanded in $\bm q$ using the Taylor expansion. The higher-order expansion of the terms in the $\bm q\to0$ limit is given as,\\ 

\begin{enumerate}
    \item Fermi-Dirac distribution function difference
    \begin{align}\nonumber
f(\varepsilon^m_{\bm {k+ q}})-f(\varepsilon^n_{\bm k}) 
&= f(\varepsilon^m_{\bm k})-f\left(\varepsilon^n_{\bm k}\right) +\sum_{\gamma} q_\gamma\,\partial_\gamma f(\varepsilon^m_{\bm k})
+\frac{1}{2}\sum_{\alpha, \beta}q_\alpha q_\beta\,\partial_{\alpha\beta} f(\varepsilon^m_{\bm k}) \\& +\frac{1}{6}\sum_{\alpha', \beta', \ell}q_{\alpha'} q_{\beta'} q_\ell\,\partial_{\alpha'\beta'\ell} f(\varepsilon^m_{\bm k})
+\mathcal {O}(q^4),
\end{align}
where $\partial_{\gamma}\equiv\partial /\partial k_{\gamma}$ and $\mathcal{O}$ refers to higher order terms.

      \item Energy dispersion difference
\begin{equation}
\varepsilon^m_{\bm k+\bm q}-\varepsilon^n_{\bm k}
= \omega^{mn}
+\sum_{\gamma} q_\gamma\,\partial_\gamma \varepsilon^m_{\bm k}
+ \frac{1}{2}\sum_{\alpha, \beta}q_\alpha q_\beta\,\partial_{\alpha\beta}\varepsilon^m_{\bm k}
+ \frac{1}{6}\sum_{\alpha', \beta', \ell}q_{\alpha'} q_{\beta'} q_\ell\,\partial_{\alpha'\beta'\ell}\varepsilon^m_{\bm k}
+ \mathcal O(q^4),
\end{equation}
where $\omega^{mn} = \varepsilon^m_{\mathbf k}-\varepsilon^n_{\mathbf k}.$

\item Angular factor $\cos\big(2\delta\theta_{\bm{kq}}^{mn} \big)$\\
Let us first consider
\begin{align}\nonumber
\delta\theta_{\bm{kq}}^{mn} &= \theta^m_{\bm{k+q}}-\theta^n_{\bm k}\\
&=\delta\theta_{\bm{k}}^{mn} + \sum_{\gamma} q_\gamma\,\partial_\gamma \theta^m_{\bm k}
+\frac{1}{2}\sum_{\alpha, \beta} q_\alpha q_\beta\,\partial_{\alpha\beta}\theta^m_{\bm k}
+ \frac{1}{6}\sum_{\alpha', \beta', \ell}q_{\alpha'} q_{\beta'} q_\ell\,\partial_{\alpha'\beta'\ell}\theta^m_{\bm k}
+ \mathcal O(q^4).
\end{align}
Here $\delta\theta_{\bm k}^{mn} = \theta^m_{\bm k}-\theta^n_{\bm k}$. On further expanding $\cos(A+h)$ for small $h$ where $A= \delta\theta_{\bm k}^{mn}$ and $h=\sum_{\gamma} q_\gamma\,\partial_\gamma \theta^m_{\bm k}
+\sum_{\alpha, \beta} \frac{1}{2}q_\alpha q_\beta\,\partial_{\alpha\beta}\theta^m_{\bm k}
+ \mathcal O(q^3)$ yields $\cos(A+h) = \cos A -h \sin A - \frac{h^2}{2}\cos A+\frac{h^3}{6} \sin A + \cdots$. Following the expansion, the angular factor becomes %Hence, the expression for the overlap matrix element becomes,
%the identity $\cos(A+h) = \cos A \cos h - \sin A \sin h$, 
%
\begin{equation}
\cos\big(2\delta\theta_{\bm{kq}}^{mn} \big)
= \cos(2\delta\theta_{\bm k}^{mn})
- 2h\sin(2\delta\theta_{\bm k}^{mn})
- 2 h^2\,\cos(2\delta\theta_{\bm k}^{mn})
+ \frac{4}{3} h^3\,\sin(2\delta\theta_{\bm k}^{mn})
+ \mathcal O(q^4).
\end{equation}
Keeping the terms up to $\mathcal O(q^2)$ we get:
\begin{align}\nonumber
\cos\big(2\delta\theta^{mn}_{\bm{kq}}\big)
&= \cos(2\delta\theta_{\bm k}^{mn})
- 2\sin(2\delta\theta_{\bm k}^{mn}) \big(\sum_{\gamma} q_\gamma\,\partial_\gamma \theta^m_{\bm k} \big) - \sin(2\delta\theta_{\bm k}^{mn})\,\big(\sum_{\alpha, \beta}q_\alpha q_\beta\,\partial_{\alpha\beta}\theta^m_{\bm k}\big) \\ &- 2\cos(2\delta\theta_{\bm k}^{mn})\big(\sum_{\gamma}q_\gamma \partial_\gamma \theta^m_{\bm k}\big)^2
+ \mathcal O(q^3).
\end{align}
\end{enumerate}
Using the expansions of all above factors. the form of the polarizability is written as
\begin{align} \label{eqn:polarizability_q->0A} \nonumber
\text{P}^{mn}(\mathbf{q}\rightarrow 0, \omega) &= \frac{1}{2} \sum_{\bm k} 
\frac{f^{mn}+ \sum_\gamma q_\gamma \partial_{\gamma} f(\varepsilon_{\bm k}^m)+\frac{1}{2}\sum_{\alpha, \beta}q_\alpha q_\beta\,\partial_{\alpha\beta} f(\varepsilon^m_{\bm k})}{\omega + \omega^{mn} + \sum _{\gamma'}q_{\gamma'}\partial_{\gamma'} \varepsilon_{\bm k}^m+  \frac{1}{2}\sum_{\alpha', \beta'}q_{\alpha'} q_{\beta'}\,\partial_{\alpha'\beta'}\varepsilon^m_{\bm k} + i\eta}\\\nonumber
&\times \big[ 1- \sum_{\gamma''} mn \sin2\delta\theta_{\bm k}^{mn}  q_{\gamma''} \partial_{\gamma''}\theta_{\bm k}^m (1+(q_{\gamma''} \partial_{\gamma''} \theta_{\bm k}^m)^2 + \cdots)
\\ & +mn\cos 2\delta\theta_{\bm k}^{mn} ( 1 - \{2 \sum_{\gamma''} q_{\gamma''} \partial_{\gamma''} \theta_{\bm k}^m\}^2+\cdots)\big].%+\frac{4}{3}\{2 \sum_\gamma q_\gamma \partial_\alpha \theta_{\bm k}^m\}^4.
\end{align}
Here we define $f^{mn} = f(\varepsilon^m_{\bm k}) - f(\varepsilon^n_{\bm k})$. This is the Eq.~\eqref{eqn:polarizability_sim} in the main text.
Specifically, for the low-energy two-band model of a Dirac nodal line semimetal considered in this work, we have $\theta^m_{\bm k}=\theta^n_{\bm k}$. Consequently $\sin 2\delta \theta_{\bm k}^{mm}=\sin 2\delta \theta_{\bm k}^{mn}=0$ and $\cos 2\delta \theta_{\bm k}^{mm} = \cos 2\delta \theta_{\bm k}^{mn}  = 1$, this simplifies Eq.~\eqref{eqn:polarizability_q->0A} to the form, 
\begin{align} \label{eqn:polarizability1_q->0} \nonumber
\text{P}^{mn}(\mathbf{q}\rightarrow 0, \omega) &
=\frac{1}{2} \sum_{\bm k} 
\frac{f^{mn}+ \sum_\gamma q_\gamma \partial_{\gamma} f(\varepsilon_{\bm k}^m)+\frac{1}{2}\sum_{\alpha, \beta}q_\alpha q_\beta\,\partial_{\alpha\beta} f(\varepsilon^m_{\bm k})}{\omega + \omega^{mn} + \sum _{\gamma'}q_{\gamma'}\partial_{\gamma'} \varepsilon_{\bm k}^m+  \frac{1}{2}\sum_{\alpha', \beta'}q_{\alpha'} q_{\beta'}\,\partial_{\alpha'\beta'}\varepsilon^m_{\bm k} + i\eta} \\& \times \big[ 1+mn ( 1 - \{2 \sum_{\gamma''} q_{\gamma''} \partial_{\gamma''} \theta_{\bm k}^m\}^2+\cdots)\big].
\end{align}
Further, in the case of intraband polarizability ($m=n$), and using the binomial expansion, the expression becomes,
\begin{align} \label{eqn:polarizability2_q->0} \nonumber
\text{P}^{mm}(\mathbf{q}\rightarrow 0, \omega)
&= \frac{1}{\omega} \sum_{\bm k} 
\Big(\sum_\gamma q_\gamma \partial_{\gamma} f(\varepsilon_{\bm k}^m)+\frac{1}{2}\sum_{\alpha, \beta}q_\alpha q_\beta\,\partial_{\alpha\beta} f(\varepsilon^m_{\bm k})\Big)\\ & \times \Big(1  - \frac{\sum _{\gamma'}q_{\gamma'}\partial_{\gamma'} \varepsilon_{\bm k}^{m}+  \frac{1}{2}\sum_{\alpha', \beta'}q_{\alpha'} q_{\beta'}\,\partial_{\alpha'\beta'}\varepsilon^{m}_{\bm k}}{\omega}\Big).
\end{align}
Here, we use $f^{mm}=0$ and $\omega^{mm} = 0$. Further using $\partial_{\gamma}f(\varepsilon_{\bm k}^m) = \frac{\partial f(\varepsilon_{\bm k}^m) }{\partial \varepsilon_{\bm k}^m} \partial_{\gamma} \varepsilon_{\bm k}^m$ and keeping the leading order term in $q$, the above expression reduces to,
\begin{align}
   \text{P}^{mm}(\mathbf{q}\rightarrow 0,\omega)
= - \sum_{\bm k}\sum_{\gamma,\gamma'}  \frac{q_\gamma q_{\gamma'}}{\omega^2} \frac{\partial \varepsilon^m_{\bm k}}{\partial k_\gamma}\frac{\partial \varepsilon^m_{\bm k}}{\partial k_{\gamma'}} \frac{\partial f(\varepsilon^m_{\bm k})}{\partial \varepsilon^m_{\bm k}}.
\end{align}
It is to be noted that the term associated with linear order $q_{\gamma}$ in the above expression does not contribute due to the odd function in $\mathcal{K}$, giving us the $q^2$ leading order dependence on the intraband polarizability. This refers to Eq.~\eqref{eqn:polarizability_low temperature} in the main text.\\
Further, in the case of interband ($m\neq n$) in Eq.~\eqref{eqn:polarizability1_q->0}, the interband polarizability becomes,
\begin{align} 
\text{P}^{mn}(\mathbf{q}\rightarrow 0, \omega) &=\frac{1}{2} \sum_{\bm k} 
\frac{f^{mn}+ \sum_\gamma q_\gamma \partial_{\gamma} f(\varepsilon_{\bm k}^m)+\frac{1}{2}\sum_{\alpha, \beta}q_\alpha q_\beta\,\partial_{\alpha\beta} f(\varepsilon^m_{\bm k})}{\omega + \omega^{mn} + \sum _{\gamma'}q_{\gamma'}\partial_{\gamma'} \varepsilon_{\bm k}^m +  \frac{1}{2}\sum_{\alpha', \beta'}q_{\alpha'} q_{\beta'}\,\partial_{\alpha'\beta'}\varepsilon^m_{\bm k} + i\eta} \\\nonumber &\times\big[ \{2 \sum_{\gamma''} q_{\gamma''} \partial_{\gamma''} \theta_{\bm k}^m\}^2-\cdots\big].
\end{align}
Keeping the above expression to the leading order in $q$, we get
\begin{align}
\text{P}^{mn}(\mathbf{q}\rightarrow 0, \omega) &=2 \sum_{\bm k}  \sum_{\gamma''}  \big(q_{\gamma''} \partial_{\gamma''} \theta_{\bm k}^m\big)^2~
\frac{f^{mn}+ \sum_\gamma q_\gamma \partial_{\gamma} f(\varepsilon_{\bm k}^m)}{\omega + \omega^{mn} + i\eta} .
\end{align}
On taking $f^{mn} = -1 +\Theta(\mu-\varepsilon_{\bm k}^{m})$ and $\partial_{\gamma}f(\varepsilon_{\bm k}^m) = \frac{\partial f(\varepsilon_{\bm k}^m) }{\partial \varepsilon_{\bm k}^m} \partial_{\gamma} \varepsilon_{\bm k}^m = -\delta(\mu - \varepsilon_{\bm k}^m) \partial_{\gamma} \varepsilon_{\bm k}^m$, we have
\begin{align} 
\label{eqn:P+-_2a_1}
 &\text{P}^{mn}(\bm q\rightarrow 0, \omega) =  2 \sum_{\bm k}\sum_{\gamma''}\big(  q_{\gamma''} \partial_{\gamma''} \theta_{\bm k}^m\big)^2~~ \frac{-1 +\Theta(\mu-\varepsilon_{\bm k}^{m})- \sum_\gamma q_\gamma \frac{\partial \varepsilon_{\bm k}^{m}}{\partial \bm k_\gamma}\delta(\mu -\varepsilon_{\bm k}^{m})}{\omega + \omega^{mn} + i\eta}.
\end{align}
This refers to Eq.~\eqref{eqn:P+-_2a} in the main text.\\
From the above analysis, it is evident that the presence of the higher order term introduces the contribution from $q^4$, $q^5$ and $q^6$ in polarizability, which can be neglected in comparison to quadratic and cubic terms in $q$. 

\section{Explicit forms of functions used in Eq.~\eqref{eqn:P_2^(+-)_case2}} 
\label{Appendix:c}
The expression for the $\text{P}^{+-}$ part of interband polarizability, as presented in Eq.~\eqref{eqn:10}

\begin{align}
\text{P}_2^{+-}=\frac{\tilde{q}_z}{24 \pi^2  (\tilde{\omega} + 2\tilde{\mu}+i\eta)(\tilde{\mu}^2 -\tilde{M}^2)^2} \left[
\tilde{q}_\rho^2 \mathcal{B}_1(\tilde{M}, \tilde{\mu})+ \gamma \tilde{q}_z \left( \tilde{q}_\rho \mathcal{B}_2(\tilde{M}, \tilde{\mu}) + \gamma \tilde{q}_z \mathcal{B}_3(\tilde{M}, \tilde{\mu})\right)
\right].
\end{align}

Here, $\mathcal{B}_1(\tilde{M}, \tilde{\mu})= 4 \tilde{\mu}   (2  \tilde{\mu}^2-3 \tilde{M}^2)$, $\mathcal{B}_2(\tilde{M}, \tilde{\mu})=  3\big[ \tilde{M} \tilde{\mu} (\tilde{M}^2 + \tilde{\mu}^2) - (\tilde{M}^2 -  \tilde{\mu}^2)^2 \tan^{-1}\big( \frac{ \tilde{\mu}}{\tilde{M}} \big) \big]$ and  $\mathcal{B}_3(\tilde{M}, \tilde{\mu})= 4 \tilde{\mu}^3$.\\

The expression for the $\text{P}^{+-}$ part of interband polarizability, as presented in Eq.~\eqref{eqn:P_2^(+-)_case2}
\begin{align}
    \text{P}_2^{+-} = \frac{\tilde{q}_z}{48 \pi^2 (\tilde{M}^2 - \tilde{\mu}^2)^2 (2\tilde{\mu} + \tilde{\omega})}
\left[
\tilde{q}_\rho^2 \mathcal{C}_1(\tilde{M}, \tilde{\mu})+ \gamma \tilde{q}_z \left( \tilde{q}_\rho \mathcal{C}_2(\tilde{M}, \tilde{\mu}) + \gamma \tilde{q}_z \mathcal{C}_3(\tilde{M}, \tilde{\mu})\right)
\right].
\end{align}
Here,
\begin{align}\nonumber
&\mathcal{C}_1(\tilde{M}, \tilde{\mu}) = 3 - 4S + \tilde{M}^2 ( 12  - 16  S - 12  \tilde{\mu}(1 + \tilde{\mu})) + \tilde{\mu}^2 \left(-6 + 12 S + \tilde{\mu}(8 + 3\tilde{\mu})\right)+ 9 \tilde{M}^4\\\nonumber
& \mathcal{C}_2(\tilde{M}, \tilde{\mu}) = 2(4 - 3S) R - \tilde{M}^2\left(8 \tilde{M} + 3 \tilde{M} \tilde{\mu} - 16 R + 9 S R\right)+ \tilde{\mu}^2 \left(-3 \tilde{M} \tilde{\mu} + (-8 + 3S) R \right) \\\nonumber
&~~~~~~~~~~~~~~~~ - 3(\tilde{M}^2 - \tilde{\mu}^2)^2 \left( \tan^{-1}\bigg[\frac{\tilde{\mu}}{\tilde{M}} \bigg] +  \tan^{-1}\bigg[\frac{S}{R} \bigg] \right)\\
& \mathcal{C}_3(\tilde{M}, \tilde{\mu}) = -3 + 4S + \tilde{M}^2(-6 + 4S) + \tilde{\mu}^3(4 + 3\tilde{\mu}) - 3\tilde{M}^4,
\end{align}
where terms $S = \sqrt{1 + \tilde{M}^2}$, and $R = \sqrt{|\tilde{\mu}^2 - 2\tilde{M}^2 - 1|}$.

\section{Explicit forms of functions used in the Eq.~\eqref{P++_1}} \label{Appendix:A}
The expression for the intraband polarizability, as presented in Eq.~\eqref{P++_1}
\begin{align} \label{P++_2}
  &  \text{P}^{++}= \frac{1}{24 \pi^2 \gamma^2 \tilde{\mu} \tilde{\omega}^2}\bigg[\mathcal{A}_1(\tilde{M}, \tilde{\mu})~\tilde{ q}_{\rho}^2 +  \mathcal{A}_2(\tilde{M}, \tilde{\mu})~\gamma^2 \tilde{q}_{z}^2\bigg],
\end{align}
where the function $\mathcal{A}_i(\tilde{M}, \tilde{\mu})$ ($i = 1,2$) is defined as:
\begin{align}%\nonumber
&\mathcal{A}_i(\tilde{M},\tilde{\mu})= \mathcal{G}(\tilde{M},\tilde{\mu}) + \tilde{M}~ a_i(\tilde{M}, \tilde{\mu})+\sqrt{|\tilde{\mu}^2 - 2\tilde{M}^2-1|}~\Delta_i(\tilde{M}, \tilde{\mu}).%\\  
\end{align}
The auxiliary functions used in the above expression are given by: 
\begin{align}\nonumber
    &\mathcal{G}(\tilde{M},\tilde{\mu}) =3 (\tilde{\mu}^2 - \tilde{M}^2)\bigg (\tan^{-1} \bigg[\frac{\tilde{\mu}}{\tilde{M}} \bigg]+ \tan^{-1} \bigg[\frac{\sqrt{1+\tilde{M}^2}}{\sqrt{|\tilde{\mu}^2 -2\tilde{M}^2 - 1|}}\bigg]\bigg)\\\nonumber
   & a_1(\tilde{M}, \tilde{\mu})= 4\tilde{M}^2 - 3\tilde{\mu}(2\tilde{\mu}+1), ~~~~~~~~~~~~~~~ \Delta_1(\tilde{M}, \tilde{\mu})= 4\tilde{\mu}^2-2\tilde{M}^2 -3\sqrt{1+\tilde{M}^2}+2\\\nonumber
   &a_2(\tilde{M}, \tilde{\mu})= 2\tilde{M}^2 + 3\tilde{\mu}, ~~~~~~~~~~~~~~~~~~~~~~~~~~ \Delta_2(\tilde{M}, \tilde{\mu})= 2\tilde{\mu}^2 - 4 \tilde{M}^2 + 3 \sqrt{1+\tilde{M}^2} - 2.
\end{align}

\section{Derivation of component of Intraband polarizability $\text{P}^{--}$} \label{Appendix:B}
The general form of polarizability is
\begin{align} \label{eqn: P--}
    &\text{P}^{--}(\bm q, \omega) = \frac{1}{2 v} \sum_{\bm k} \frac{f( \varepsilon^-_{\bm k+\bm q})-f(\varepsilon^-_{\bm k})}{\omega+ \varepsilon^-_{\bm k+\bm q}-\varepsilon^-_{\bm k}+i \eta} |\langle m, \bm k +\bm q|n, \bm k\rangle|^2.
\end{align}
In the low-temperature limit, the Fermi Dirac distribution function reduces to the Heaviside theta function, i.e., $f(\varepsilon_{\bm k})=\Theta (\mu - \varepsilon_{\bm k})$, hence the above equation becomes, 

\begin{align} \label{eqn: P--}
    &\text{P}^{--}(\bm q, \omega) = \frac{1}{2 v} \sum_{\bm k} \frac{\Theta (\mu + \varepsilon_{\bm k+\bm q})-\Theta (\mu + \varepsilon_{\bm k})}{\omega- \varepsilon_{\bm k+\bm q}+\varepsilon_{\bm k}+i \eta} \times (1+\cos{2 \delta \theta_{\bm{kq}}})
\end{align}

The Heaviside theta function is defined as $\Theta(x)=0$ for $x<0$ and $\Theta(x)=1$ for $x>0$. In the above expression, $\Theta(\mu + \varepsilon_{\bm k})=1$ when $\mu + \varepsilon_{\bm k}>0$, which satisfies conditions $\varepsilon_{\bm k}>0$ and $\mu>0$. Under this condition, both $\Theta (\mu + \varepsilon_{\bm k+\bm q})$ and $\Theta (\mu + \varepsilon_{\bm k})$ become $1$,  and the term $\Theta (\mu + \varepsilon_{\bm k+\bm q})-\Theta (\mu + \varepsilon_{\bm k})= 1-1=0$. As a result, the $\text{P}^{--}$ contribution to the intraband polarizability vanishes. 
In addition, when $\Theta(\mu + \varepsilon_{\bm k})=0$ for $\mu + \varepsilon_{\bm k}<0$, the step function becomes zero, hence $\text{P}^{--}$ zero.  
\twocolumngrid
\bibliography{Ref}

\end{document}